\documentclass[smallextended]{svjour3}
\smartqed  % flush right qed marks, e.g. at end of proof
\usepackage{graphicx}
\usepackage{epstopdf}
\DeclareGraphicsExtensions{.png,.pdf}
\usepackage[nomarkers,figuresonly]{endfloat}
\usepackage{float}
\usepackage[caption = false]{subfig}
\usepackage{natbib}
\usepackage{amssymb}
\usepackage{rotating}
\usepackage{url}

\newcommand{\apj}{ApJ}
\newcommand{\apss}{A\&SS}

\newcommand{\apjl}{ApJL}
\newcommand{\mnras}{MNRAS}
\newcommand{\apjs}{ApJS}
\newcommand{\pasp}{PASP}
\newcommand{\araa}{ARA\&A}
\newcommand{\aap}{A\&A}
\newcommand{\aaps}{A\&AS}

\newcommand{\bain}{BAIN}
\newcommand{\nat}{Nature}
\newcommand\ion[2]{#1$\;${\scshape{#2}}}% 

\begin{document}

%\title{Interstellar H$_2$ in the $\epsilon$ Persei Sightline: Results from the Colorado High Resolution Echelle Spectrograph (CHESS)}

%\title{Interstellar H$_2$ in the $\epsilon$ Persei Sightline: Results from the Colorado High Resolution Echelle Spectrograph (CHESS)}

\title{CHESS: an Innovative Concept for High-Resolution, Far-UV Spectroscopy}
\subtitle{Instrument Design, Inception, and Results from the First Two Sounding Rocket Flights}
%\shorttitle{CHESS: the Far-UV Spectrograph}

%% Authors, affiliations.
%\correspondingauthor{Keri Hoadley}
%\email{khoadley@caltech.edu}
%Nicholas Kruczek \and
% \and Nicholas Kruczek 

\author{Keri Hoadley \and 
        Kevin France \and
        Nicholas Nell \and
        Robert Kane \and
        Brian Fleming \and
        Allison Youngblood \and
        Matthew Beasley
        }
\institute{Keri Hoadley \at 
            California Institute of Technology, Dept. of Physics, Mathematics, \& Astronomy, Cahill Center for Astronomy \& Astrophysics, Pasadena, CA 91125, USA \\
            David \& Ellen Lee Postdoctoral Fellow in Experimental Physics at Caltech \\
			University of Iowa, Dept. of Physics \& Astronomy, Van Allen Hall, Iowa City, IA 52242 USA \\
                \email{khoadley@caltech.edu} \\
                %\orcid{0000-0002-8636-3309}
             \and
            Kevin France \and Nicholas Nell \and Brian Fleming \and Allison Youngblood \at
                Laboratory for Atmospheric and Space Physics, University of Colorado, UCB 392, Boulder, CO 80309, USA 
             \and
            Robert Kane \at 
                %SEAKR Engineering, 6221 S. Racine Circle, Centennial, CO 80111, USA
                Blue Canyon Technologies, 2550 Crescent Drive, Lafayette, CO 80026, USA
             %\and
            %Allison Youngblood \at 
                %Exoplanets \& Stellar Astrophysics Laboratory, NASA Goddard Space Flight Center, Code 667, Greenbelt, MD 20771, USA
             \and
            Matthew Beasley \at 
                Southwest Research Institute, Department of Space Studies, 1050 Walnut Ave., Boulder, CO, 80302, USA
          }

\date{Received: date / Accepted: date}

\maketitle

%% Abstract section.
\begin{abstract}
 The space ultraviolet (UV) is a critical astronomical observing window, where a multitude of atomic, ionic, and molecular signatures provide crucial insight into planetary, interstellar, stellar, intergalactic, and extragalactic objects. The next generation of large space telescopes require highly sensitive, moderate-to-high resolution UV spectrograph. However, sensitive observations in the UV are difficult, as UV optical performance and imaging efficiencies have lagged behind counterparts in the visible and infrared regimes. This has historically resulted in simple, low-bounce instruments to increase sensitivity.
 In this study, we present the design, fabrication, and calibration of a simple, high resolution, high throughput FUV spectrograph - the \emph{Colorado High-resolution Echelle Stellar Spectrograph} (CHESS). CHESS is a sounding rocket payload to demonstrate the instrument design for the next-generation UV space telescopes. 
 We present tests and results on the performance of several state-of-the-art diffraction grating and detector technologies for FUV astronomical applications that were flown aboard the first two iterations of CHESS. The CHESS spectrograph was used to study the atomic-to-molecular transitions within translucent cloud regions in the interstellar medium (ISM) through absorption spectroscopy. The first two flights looked at the sightlines towards $\alpha$ Virgo and $\epsilon$ Persei and flight results are presented.
 % The signal-to-noise ratio of the $\alpha$ Virgo observation was too low to obtain meaningful results. However, several key pieces of information about the $\epsilon$ Persei sightline were extracted from the second flight observations, which are presented. We report on the presence of metals and ions in the sightline, and find that the molecular hydrogen sightline is consistent with previous studies performed with \emph{IUE} and \emph{Copernicus}.
 
 %During the alignment and calibration of the first iterations of CHESS, it was discovered that manufacturing errors prevented the instrument from performing at its full potential. This study reports on the initial performance of CHESS and its first two demonstration flights. Subsequent 

\end{abstract}

%% Keywords: Always after the abstract.

\section{Introduction}

Ultraviolet (UV) observations are critical for addressing a many key questions in nearly all aspects of astrophysics. UV observations probe the flow of energetics in the universe as UV photons affect atoms, molecules, ions, and dust. The UV is the peak of the hot star spectral energy distribution (SED), making it the ideal probe of massive (recent) star-formation and galactic star formation histories (e.g., %one of the primary science objectives of the {\it Galaxy Evolution Explorer} ({\it GALEX}; 
\citealt{Martin+2005}). A multitude of atomic, ionic, and molecular species have strong resonances with UV radiation, which pump and produce strong emission and absorption lines seen against the UV background. The UV is home to emission lines in collisional ionization equilibrium at formation temperatures up to $\sim$3$\times$10$^{5}$ K \citep{Tumlinson+2017}, critical for assessing hot interstellar, circumgalactic, and intergalactic environments. The only other wavelength regime with the potential to observe so many phases of gas and dust simultaneously is the far-IR, which similarly requires access to space. 
In cooler regimes of the interstellar medium (ISM), dust scatters and absorbs far-UV (FUV) photons, which provides a UV background that changes the chemistry of galactic ISMs (e.g., \citealt{Hamden+13}). While dust is often seen as a hindrance to UV observations, such sightlines can be used to sensitively measure fine dust chemistry and distributions in different astrophysical environments (e.g., \citealt{Blasberger+2017,Ma+2020}). Similarly, UV spectroscopy of relatively high reddedning (E(B-V) $>$ 0.5) sightlines have routinely been observed from space since the 1970s, with the {\it International Ultraviolet Explorer} ({\it IUE}) and Copernicus revolutionizing our understanding of the diffuse and translucent ISM physical state and chemistry using UV absorption line spectroscopy (e.g., \citealt{Spitzer+73,Morton+75,Savage+77,Bohlin+83}, and many others). The UV is the only wavelength regime we gain access to cold molecular hydrogen (H$_2$; T(H$_2$) $<$ 500 K), and this in itself is a critical diagnostic for many astrophysical environments and properties that rely on accurate measures of H$_2$ density or mass. %, yet often assume how much H$_2$ is there based on trends or ratios with other tracer species (e.g, $X_{CO}$). 
UV observations provide a critical look at the characteristics of many astrophysical systems, both nearby and afar.

Present-day ultraviolet spectroscopic facilities on community space telescopes (currently, only the {\it Hubble Space Telescope} ({\it HST}) Cosmic Origins Spectrograph (COS; \citealt{Green+12}) and Space Telescope Imaging Spectrograph (STIS; \citealt{Kimble+1998}) provide spectroscopic coverage in the FUV) have provided a unique view of many astrophysical systems, but they have their limitations and cannot address all our remaining questions. Current space facility spectrographs rely on old technologies in reflective mirror coatings, gratings, and detectors that all perform moderately in the UV - many gains in improvement can be made in almost all realms of UV instrumentation to probe deeper and fainter objects in our universe. Many of the most interesting, intriguing, and mysterious diffuse objects discovered by the \emph{Galaxy Evolution Explorer} (\emph{GALEX}) a decade ago (e.g., \citealt{Martin+2007,Sahai2010,Hoadley+2020}) are unattainable with UV spectrographs on \emph{HST}, which do not have the sensitivity to observe them. 
%, but 
Highly sensitive, high-resolution UV spectroscopy is necessary to address astronomical questions that are beyond the reach of present-day instruments. In this paper, we discuss a concept for a high-throughput, high-resolution (R $>$ 100,000) far-UV (1000 - 1600 {\AA}) spectrograph to serve as an instrument baseline for the next generation of UV-Visible space telescopes (e.g, \citealt{France+16,Woodruff+2019,luvoir2019}). 

\subsection{Instrument Overview}

The \emph{Colorado High-resolution Echelle Stellar Spectrograph} (CHESS) is an objective echelle spectrograph designed to achieve a minimum resolving power (R) $>$ 100,000 through the far-UV (1000 $-$ 1600 {\AA}) \citep{Hoadley+16,Hoadley+14,France+12,Kane+11,Beasley+10,France+16JAI}. CHESS does not utilize a telescope, but rather light is fed into the spectrograph through a collimator. In this way, the instrument consists of a 2-bounce system, minimizing the number of reflections required of the far-UV photons before being captured by the detector. The focusing optic of the system is tied into the cross-dispersing grating, which also includes additional surface curvature orthogonal to the focusing axis to correct for first-order aberrations in the spectral line. %: the first launch aboard NASA/CU mission 36.285 UG on 24 May 2014 (CHESS-1), and the second flight aboard the NASA/CU mission 36.297 UG on 21 February 2016 (CHESS-2). Both CHESS launches took place at White Sands Missile Range (WSMR) in southern New Mexico.

CHESS demonstrates an innovative high-resolution FUV spectrograph design, ideal as an ancillary instrument for future space telescopes. The instrument also serves as a platform to test, verify, and flight-qualify the performance of both grating and detector technologies for UV astrophysical applications. Highly sensitive instruments for future missions will require high efficiency and low scattered light and background levels, which are of the highest priority to address for the UV instrumentation community (e.g. \citealt{Scowen+17} and references therein). All optical (echelle and cross-dispersing gratings) and detector (cross-strip anode microchannel plate (MCP)) components on CHESS were experimental, with technology readiness levels (TRLs) $\leq$ 6. 

CHESS is a rocket-borne astronomical instrument that has launched on four separate NASA-funded sounding rocket missions. We report on the design, build, calibration, and integration leading to the first two flights of CHESS. We present the observations from both launches, and an in-depth analysis of the sightline observed on the second flight.

%The spectrograph was designed to employ an R2 echelle grating with a low line density ($<$ 100 lines/mm), using higher-order diffraction solutions to achieve R $>$ 100,000 throughout the instrument bandpass. The cross-dispersing grating, developed and ruled by Horiba Jobin-Yvon, is a holographically-ruled, ``low" line density, powered optic with a toroidal surface curvature. Both gratings were coated with aluminum and lithium fluoride (Al+LiF) at Goddard Space Flight Center (GSFC). Results from final efficiency and reflectivity measurements for the optical components of CHESS-2 are presented. CHESS-2 utilizes a 40mm-diameter cross-strip anode readout microchannel plate (MCP) detector fabricated by Sensor Sciences, Inc., to achieve high spatial resolution with high count rate capabilities (global rates $\sim$ 1 MHz). 

\subsection{Scientific Program}

The FUV bandpass and high spectral resolution of CHESS are ideally suited to study interstellar structures in the sightlines of hot, massive O- and B-type stars.

%\subsubsection{Translucent Clouds}
\textbf{Translucent clouds} reside in the transition between the diffuse (traditionally defined as the visual dust extinction (A$_{V}$) $<$ 1) and dense (A$_{V}$ $>$ 3) phases of the interstellar medium (ISM). It is in this regime where the UV portion of the average interstellar radiation field plays a critical role in the photochemistry of the gas and dust clouds that pervade the Milky Way galaxy. One powerful technique for probing the chemical structure of translucent clouds is to combine measurements of H$_{2}$ with knowledge of the full carbon inventory (CI, CII, and carbon monoxide (CO)) along a given line of sight. \citet{Snow+06b} argue that an analysis of the carbon budget should be the defining criterion for translucent clouds, rather than simple measurements of visual extinction. Moderate resolution 1000-1120 {\AA} spectra from FUSE and higher-resolution spectra from HST/STIS can confirm whether or not a sightline is consistent with the existence of translucent material in the framework of current models of photodissociation regions of the ISM (CO/H$_2$ $>$ 10$^{-6}$ and CO/CI $\sim$ 1; \citealt{Burgh+07,Burgh+10}).
%It has been argued that an analysis of the carbon budget should be the defining criterion for translucent clouds, rather than simple measurements of visual extinction \citep{Snow+06b}. Moderate resolution 1000 $-$ 1120 {\AA} spectra from FUSE and higher-resolution spectra from HST/STIS have been used to show that many of these sightlines have CO/H$_{2}$ $>$ 10$^{-6}$ and CO/CI $\sim$ 1, consistent with the existence of translucent material in the framework of current models of photodissociation regions in the ISM \citep{Burgh+07,Burgh+10}.

The bandpass of CHESS contains absorption lines of H$_{2}$ (1000 $-$ 1120 {\AA}), CII (1036 and 1335 {\AA}), CI (1103 $-$ 1130, 1261, 1561 {\AA}), and several bands of CO ($<$ 1510 {\AA}). High resolution (R $>$ 100,000) is required to resolve the velocity structure of both the CI lines and the rotational structure of CO, essential to accurately determine the column density of these species \citep{Jenkins+01}. The FUV bandpass also provides access to many ionic absorption lines to explore the depletion patterns of metals in translucent clouds. %CHESS, with its high-resolution and large bandpass, especially including wavelengths shorter than 1150 {\AA}, is well-suited to the study of translucent clouds and will help create an observational base for models of the chemistry and physical conditions in interstellar clouds.

%\subsubsection{Properties of the Local Interstellar Medium}
 %The CHESS design also provides an ideal testbed to characterize the structure and thermal properties of the local interstellar medium (LISM).
\textbf{The Local Interstellar Medium (LISM)} provides an opportunity to study general ISM phenomena up close and in three dimensions, including interactions of different phases of the ISM, cloud collisions, cloud evolution, ionization structure, thermal balance, and turbulent motions \citep{Redfield+06}. Our immediate interstellar environment also determines the structure of the heliosphere, or the momentum balance of the solar wind and the surrounding ISM. %Additionally, multiple launches of CHESS allows for multiple lines of sight through the LISM, making it possible to construct a three-dimensional morphological and physical model of the LISM. 
Several physical characteristics of the LISM are measurable, including the ionization structure. Since many clouds in the LISM are optically thin, the distribution of ionizing sources (i.e., hot stars) determines the three-dimensional ionization structure of the LISM. 

Measurements of different ionization species are required to probe different phases of the LISM. In addition to local ionization structure, temperature and elemental depletion structure are also critical to understanding the three-dimensional morphology of the LISM. The temperature distribution of the LISM can place constraints on models of the evolution of the local solar neighborhood. Determining these temperatures requires high spectral resolution so that contributions from thermal and turbulent motions can be distinguished, a capability that is achievable with CHESS.

\section{Instrument Concept}

CHESS is an objective f/12.4 echelle spectrograph. The instrument design included the development of two novel grating technologies and flight-testing of a cross-strip anode microchannel plate (MCP) detector \citep{Beasley+10}. The high-resolution instrument is capable of achieving resolving powers $\ge$ 100,000 $\lambda$/$\Delta \lambda$ across a bandpass of 1000 $-$ 1600 {\AA} \citep{Hoadley+14,France+16JAI}.

\subsection{Spectrograph Design}
% ZEMAX ray trace
\begin{figure} %[ht]
   \begin{center} 
   \includegraphics[width=\textwidth]{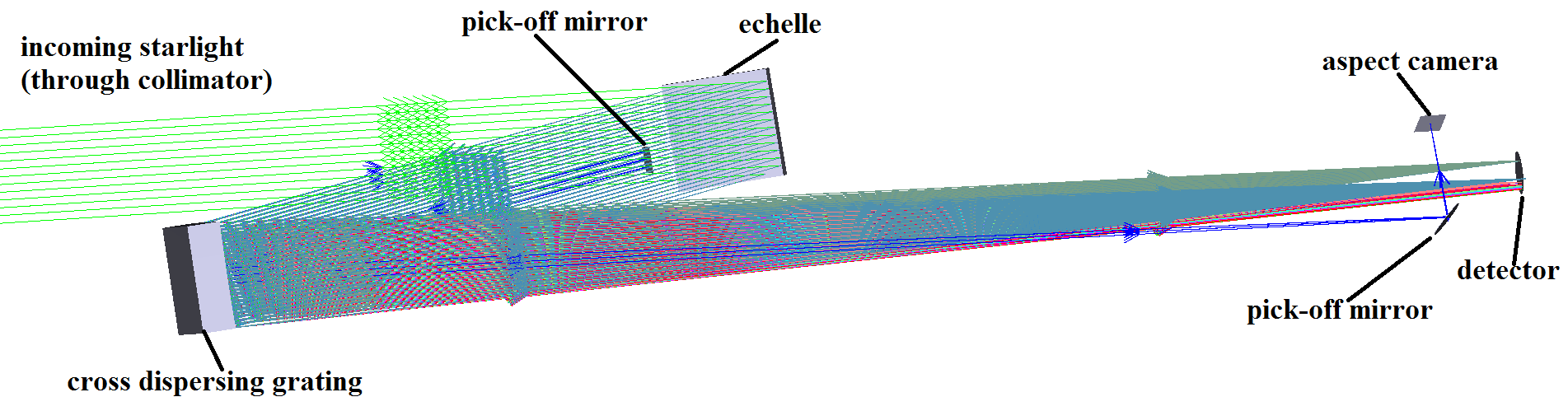}
   %{chess_raytrace_fig1.png}
   \end{center}
   \caption{The Zemax ray trace of CHESS, including the secondary aspect camera system.  The mechanical collimator reduces stray light in the line of sight and feeds starlight to the echelle. The echelle disperses UV light into high-dispersion orders, which are focused by the cross disperser onto the detector plane. The different colored lines represent a series of wavelengths across the 1000 $-$ 1600 {\AA} bandpass. Adapted from \citet{Hoadley+16}.}\label{fig:fig1} 
\end{figure}

The optical design of CHESS is as follows:

\begin{itemize}%[label=(\alph*)]

\item A mechanical collimator, consisting of an array of 10.74 mm $\times$ 10.74 mm $\times$ 1000 mm anodized aluminum tubes, provides CHESS with a total collecting area of 40 cm$^{2}$, a field of view (FOV) of 0.67$^{\circ}$, and allows on-axis stellar light through to the spectrograph. 
A mechanical collimator design was chosen over a traditional telescope design for practicality, while still achieving the necessary sensitivity for the spectrograph. A collimating telescope would be difficult to align to the spectrograph and remain aligned during the course of its flight. A wire mesh collimator was considered, but the total loss of collecting area was too high to implement. The large FOV of the mechanical collimator was dictated by the physical dimensions of commercially-available anodized aluminum tubes. Because CHESS looks at bright FUV point sources (stars) to provide a ``backlight’’ which illuminates the interstellar gas in the sightline, there was very little chance of source confusion over this FOV, given the effective area of the spectrograph and the brightness of the target stars. 

\item A square echelle grating (ruled area: 100 mm $\times$ 100 mm), with a designed groove density of 69 grooves/mm and angle of incidence (AOI) of 67$^{\circ}$, intercepts and disperses the FUV stellar light into higher diffraction terms (m = 266 $-$ 166). The custom echelle was used to research new etching technologies, namely the electron-beam (e-beam) etching process. The results of the JPL fabrication effort are discussed in Section~\ref{sec:ech}.

\item Instead of using an off-axis parabolic cross disperser \citep{Jenkins+88}, CHESS employs a holographically-ruled cross dispersing grating with a toroidal surface figure and ion-etched grooves, maximizing first-order efficiency. The cross disperser is ruled over a square area (100 mm $\times$ 100 mm) with a groove density of 351 grooves/mm and has a surface radius of curvature (RC) = 2500.25 mm and a rotation curvature ($\rho$) = 2467.96 mm. The grating spectrally disperses the echelle orders and corrects for grating aberrations \citep{Thomas+03}.

\item The cross-strip MCP detector \citep{Vallerga+10,Siegmund+09} is circular in format, 40 mm in diameter, and capable of total global count rates $\sim$ 10$^{6}$ counts/second. The cross-strip anode allows for high resolution imaging, with the location of a photoelectron cloud determined by the centroid of current read out from five anode ``fingers" along the x and y axes.

\end{itemize}

Because the echelle disperses light into high orders and the cross disperser separates light sharing the same echelle diffraction order solutions, the final data product is a series of spectra, where each echelle spectrum provides a small fraction of the total spectral coverage of the instrument. At the same time, each spectral snippet in the full raw data is able to be sampled at high resolution. This is how CHESS is able to achieve both large wavelength coverage and high resolution.

The CHESS instrument also includes an optical system (Xybion Model ISS-255 low-light video camera\footnote{At the time of CHESS-1 and 2, the Xybion Electronics System (XES) low-light camera was deprecated and the NASA Sounding Rocket Operations Contract (NSROC) was in the process of replacing it with a newer model of space-qualified low-light camera system. In 2018, NSROC officially replaced the XES Model ISS-255 with the Stanford Photonics Inc. XR/M model: \url{https://www.stanfordphotonics.com/Products/products.htm}}), which was used to align the spectrograph to an independent star tracking system during calibrations and flight. The aspect camera relies on the positions of the instrument gratings to direct zeroth-order light to a visible-light camera. Fig.~\ref{fig:fig1} shows a Zemax ray trace of collimated (star) light through the entire spectrograph. 

% Instrument layout: Mechanical
\begin{figure}%[htbp]
	\centering
	%\vspace{-5mm}
	\includegraphics[width=1.05\textwidth]{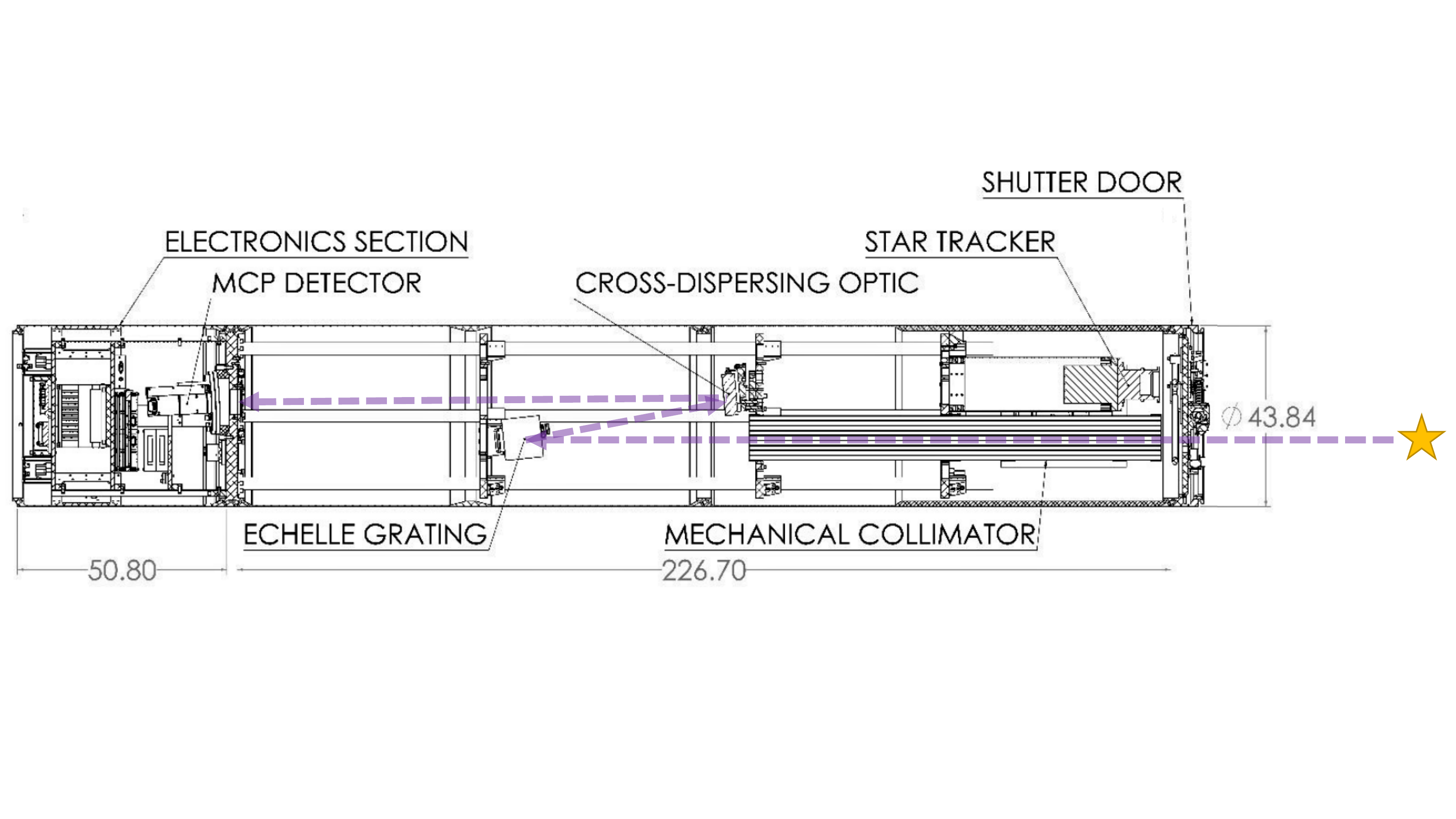}%.png}%, %resolution=3000]{f1.pdf}%{chess_mechanical_v2.eps} %, trim=0cm 0cm 0cm 6cm, angle=90
	%\vspace{-5mm}
	\caption{Schematic view of the Colorado High-resolution Echelle Stellar Spectrograph (CHESS). Dimensions defined are in centimeters. The optical path (dashed purple line) follows right to left, with the target light entering the instrument when the shutter door is open during flight.  Adapted from Adapted from \citet{Hoadley+14} and \citet{Hoadley+16}.} 
	\label{fig:chess_mech}
\end{figure}

CHESS was separated into two sections to fly aboard a sounding rocket mission: a vacuum (spectrograph) and non-vacuum (electronics) section. The two sections are separated by a hermetic bulkhead. The detector was mounted with a hermetic seal on the electronics side of the vacuum bulkhead, facing into the spectrograph section. The overall length of the payload was 226.70 cm from mating surface to mating surface, with a total weight of about 365 lbs. The opto-mechanical design of the spectrograph consisted of a carbon-fiber space frame attached to the vacuum side of the hermetic bulkhead. The carbon fiber frame (five 2.54 cm diameter x 182.88 cm long tubes) holds three light-weighted aluminum disks which suspend the mechanical collimator, echelle grating and cross-disperser in place. The aspect camera is attached to the vacuum side of the hermetic bulkhead. A Solidworks rendering of the payload is shown in Figure~\ref{fig:chess_mech}.

%%%%%%% CHESS Instrument Specifications
%
\begin{table}
\caption{Instrument Specifications for the CHESS Sounding Rocket Payload} 
\label{tab:overview}
\begin{center}       
\begin{tabular}{llll} 
\hline  \hline
	Mechanical Collimator		&		&	Spectrograph	&		\\
\hline
FOV: 18.5$^{\prime}$ $\times$ 18.5$^{\prime}$	&		&	Bandpass ({\AA}): 1000 - 1650 	&	\\
Dimensions (mm):				&		&	Resolving power:		&		\\
\qquad	10.74 $\times$ 10.74 $\times$ 1000		&		& \qquad	Theoretical R $\sim$ 120,000	&		\\
Collecting area (cm$^2$): 40.0	&		& \qquad	Demonstrated R $\leq$ 10,000	&		\\
						        &		& $F\#$: $f$/12.4			&	\\
\hline 

Echelle (CHESS-1)	& Echelle (CHESS-2)	& Cross Disperser 	& Detector	\\

\hline 
Vendor: LightSmyth	& Vendor: Bach Research	& Vendor: HORIBA Jobin-Yvon	& Vendor: Sensor Sciences \\
Shape: Flat		& Shape: Flat			& Shape: Toroidal			& Type: Open-face MCP	\\
Blaze angle($^{\circ}$): 73.0	& Blaze angle($^{\circ}$): 64.3	&  Radius (mm): 2500.25/2467.96	& Pixel format: 8k $\times$ 8k	\\
Groove density (gr/mm): 	& Groove density (gr/mm): & Groove density (gr/mm):	& Spatial resolution ($\mu$m): \\
\qquad	71.7		& \qquad	53.85			& \qquad	351			& \qquad	25			\\
Ruling: Lithographic		& Ruling: Mechanical		& Ruling: Holographic		& Anode: Cross-strip	\\
Coating: Al+LiF			& Coating: Al+LiF		& Coating: Al+LiF		& Photocathode: CsI	\\
Dimensions (mm): 		& Dimensions (mm): 		& Dimensions (mm): 		& Outer dimension (mm): 40	\\
\qquad 100 $\times$ 100 $\times$ 0.7 & \qquad 104 $\times$ 104 $\times$ 16	& \qquad 100 $\times$ 100 $\times$ 30 & Global count rate (Hz): 10$^6$ \\
Material: Silicon		& Material: Zerodur		& Material: Fused Silica	&       \\ %Material (MCPs): Borosilicate \\

\hline  \hline

\end{tabular}
\end{center}
\end{table}

\subsection{Components \& Performance}

The CHESS instrument is comprised of three optical components: two gratings (one high-dispersion echelle grating and one cross-dispersing grating) and one detector (cross-strip anode MCP). Table~\ref{tab:overview} presents technical details about each optical element used for the first two flights of the instrument.

\subsubsection{Echelle Grating Development for FUV Applications}\label{sec:ech}
Echelle gratings are distinguishable by their low line densities (20 - 300 lines/mm) and use at steep facet angles ($\theta$: 20$^{\circ}$ - 80$^{\circ}$). Both qualities allow echelle gratings to theoretically achieve high dispersion, high efficiency at or near the Littrow configuration, or where the angle of incidence equals the diffraction angle ($\alpha$ $=$ $\beta$ $=$ $\theta$), and high resolution with low polarization effects.

As a part of the instrument design, the echelle was meant to be an experimental technology demonstration piece for two different grating fabrication processes: the first was a lithographic-ruling process, provided by LightSmyth, Inc., and the second an electron-beam etching technique, fabricated by the Microdevices Laboratory at JPL. The lithographic ruling process starts with a substrate (silicon) with a thin oxidation layer, over which a photoresist and photomask with the desired groove pattern is overlaid. Using extreme-UV light to etch into the photoresist, the photomask is removed and the oxide layer is further etched via chemical agents that do not harm the photoresist. After removing the rest of the photoresist, the etched substrate is left with the desired groove density and thickness. The lithography process theoretically allows for the creation of uniform, low-scatter gratings at arbitrary groove densities with sub-100 nm surface deviations. The electron-beam etch scans a focused beam of electrons across the surface of the optic, which is covered with an electron sensitive film. The electron beam changes the solubility of the resist, which removes exposed regions of the resist in a solvent \citep{McCord+00}. This technique also enables controlled line spacing on the grating and sub-10 nm surface deviations for low scatter grooves.

\begin{figure}% [ht]
   \begin{center}
   \includegraphics[width=0.75\textwidth]{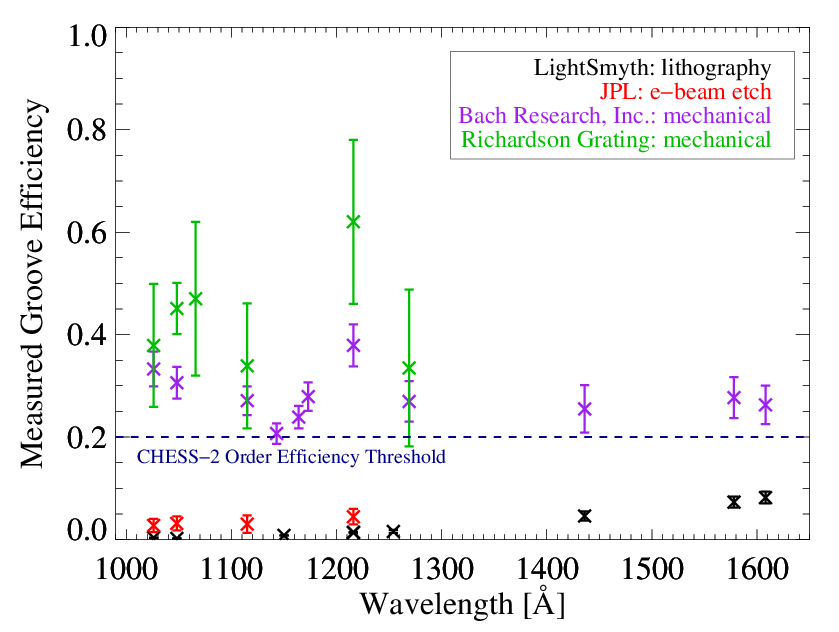}
   \end{center}
   \caption{A comparison of echelle gratings tested for use in the CHESS instrument. We include the best-performing echelle gratings from the lithography etching R\&D project undertaken by LightSmyth, Inc. (flown on CHESS-1, 36.285 UG), the e-beam samples fabricated by JPL, and two mechanically-ruled replica gratings from Bach Research, Inc. and Richardson Gratings, respectively. Both mechanically-ruled gratings out-performed the R\&D echelles and met the CHESS minimum order efficiency threshold. Adapted from \citet{Hoadley+16}.} \label{fig:ech}
\end{figure}

We present results from in-house determination of the groove efficiency of the best echelle gratings fabricated by both the lithography and electron-beam etching processes in Table~\ref{tab:ech} and Figure~\ref{fig:ech}. The high angle of incidence and high diffraction orders we were working with in CHESS made the the fabrication process of the echelle gratings extremely difficult, and both manufacturers were unable to produce echelle gratings that met our flight specification in the limited time allowed before scheduled launches. For the first flight of CHESS, we had not identified a back-up grating to use for flight, so we flew the lithographically-ruled echelle. For CHESS-2, we identified two mechanical-ruling manufacturers that provided higher efficiency gratings for the second flight. The Bach echelle was delivered in time for the second flight, while the Richardson echelle was used for subsequent flights \citep{Kruczek+17b,Kruczek+18}.

%
%
% Place Table and figure of Echelle performance here 
%
%
\begin{table}%[ht]
\caption{Comparison of Echelle Performance for CHESS \citep{Hoadley+16}} 
\label{tab:ech}
\begin{center}       
\begin{tabular}{|l|c|c|c|} 
\hline
\rule[-1ex]{0pt}{3.5ex} Grating & $\alpha$ & Groove Density & Ly$\alpha$ Efficiency \\
\rule[-1ex]{0pt}{3.5ex}  & (degrees) & (grooves/mm) &  \\
\hline
\rule[-1ex]{0pt}{3.5ex} \textbf{CHESS, Designed} & \textbf{67.0} & \textbf{69.0} & \textbf{70.0 \%}  \\
%\hline
\rule[-1ex]{0pt}{3.5ex} LightSmyth, CHESS-1 & 73.0 & 71.7 & 1.5 \%  \\
\rule[-1ex]{0pt}{3.5ex} JPL Echelle & 65.5 & 100.0 & 4.5 \% \\
%\hline
\rule[-1ex]{0pt}{3.5ex} Bach, CHESS-2$^{\star}$ & 64.3 & 53.85 & 11.9 \% + 8.4 \% \\
%\hline
%\rule[-1ex]{0pt}{3.5ex}  Richardson Echelle \#1$^{\star}$ & 64.8 & 79.0 & 30.0 \% \\
%\hline
\rule[-1ex]{0pt}{3.5ex} Richardson$^{\star}$ & 63.0 & 87.0 & 62.0 \% \\
\hline 

%\multicolumn{3}{l}{$^{\star}$ Groove densities were not specified.}
\end{tabular}
\end{center}
\end{table}

\subsubsection{Cross-Dispersing Grating}

The CHESS cross disperser grating is a 100 mm $\times$ 100 mm $\times$ 30 mm fused silica optic with a toroidal surface profile. The toroidal surface shape separates the foci of the tangential and sagittal axes of the dispersed light, which corrects astigmatic aberrations typically introduced by a more traditional off-axis parabolic design (e.g., \citealt{Sasian+1996, Jenkins+1996, Indebetouw+2001}). The surface figure of the toroid focuses the echelle order widths at the position as the spectral line widths, which are dictated by the grating solutions etched into the cross disperser. Both the sagittal and tangential foci of the toroidal optic do not intersect either the ion repeller or quantum efficiency (QE) grids, both of which are placed in front of the detector. %{\bf \st{The optic first focuses light spatially onto the detector, then spectrally behind the detector, ensuring no foci at the locations of either the ion repeller or quantum efficiency (QE) grids.}} 
The cross dispersing optic is a novel type of imaging grating that represents a new family of holographic solutions and was fabricated by Horiba Jobin-Yvon (JY). The line densities are low (351 lines per mm, difficult to achieve with the ion-etching process), and the holographic solution allows for more degrees of freedom than were previously available with off-axis parabolic cross dispersing optics. The holographic ruling corrects for aberrations that otherwise could not be corrected via mechanical ruling. The grating is developed under the formalism of toroidal variable line spacing (VLS) gratings \citep{Thomas+03} and corresponds to a holographic grating produced with an aberrated wavefront via deformable mirror technology. This results in a radial change in groove density and a traditional surface of concentric hyperboloids from holography (e.g., HST/COS; \citealt{Green+12}). 

% Figure 7 - Comparison of the performance of the cross disperser before/after 36.285 and before 36.297.
% 
\begin{figure}% [ht]
   \begin{center}
   \includegraphics[width=0.75\textwidth]{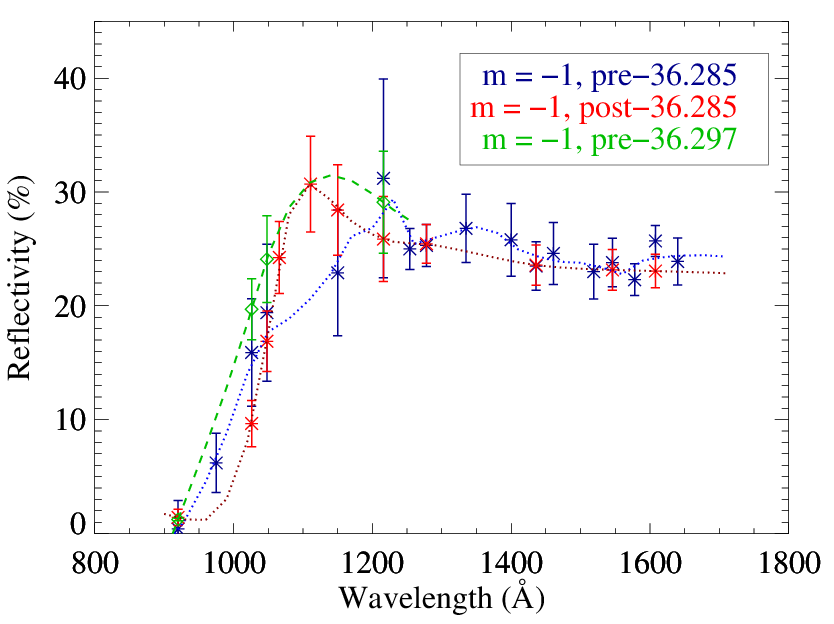}
   \end{center}
   \caption{The measured reflectivity (order efficiency $\times$ reflectivity of Al+LiF) of the cross dispersing grating in CHESS over time, overplotted with simple spline curves to show the resemblance of each trial. The colored points and lines represent times before and after the CHESS-1 and CHESS-2 missions that the cross disperser grating efficiency was measured: “pre-36.285” (blue) is before the cross disperser was installed and aligned into the instrument for the CHESS-1 flight, “post-36.285” (red) is right after we recovered the instrument after the CHESS-1 flight and measured the post-flight efficiency of all the optical components, and “pre-36.297” (green) is before the grating was installed and aligned into the instrument structure for the CHESS-2 flight. We focus on the reflectivity of the m = -1 order, which is the dispersion order used in the CHESS instrument. Because Al+LiF can exhibit efficiency degradations when not stored in a dry environment, we measure how the order reflectivity changes between CHESS-1 and CHESS-2 without re-coating the optic. No significant degradation of the coating has been measured between the first two flights of CHESS. Adapted from \citet{Hoadley+16}.} \label{fig:x-disp} 
\end{figure}

%We measured the m = +1 and m = -1 orders and found $\sim$20\% $-$ 45\% groove efficiency in the FUV (900 $-$ 1700 {\AA}) before and after the Al+LiF optical coating. 
Figure~\ref{fig:x-disp} shows the measured reflectivity (order efficiency times the reflectivity of the optical coating) of the cross dispersing optic for order m = -1, which is the dispersion order used in the CHESS instrument. We measure the reflectivity of the cross disperser before the launch of CHESS-1, between the launch of CHESS-1 and CHESS-2, and after the launch of CHESS-2, and found that it did not change significantly over time. %Overall, the performance of the cross disperser exceeded our initial expectations, with reflectivity $\sim$ 30\% at Ly$\alpha$. 
The cross disperser is effective at dispersing most of the on-axis light into the m = $\pm$ 1 orders and suppressing the m = 0 order because of the characteristic sinusoidal groove profiles created via the ion-etching procedure. Additionally, at optical wavelengths, the reflectivity of the m = 0 order becomes comparable to the m = $\pm$ 1 orders. This allowed us to build a secondary camera system to track the movements of our optical axis and target acquisition during flight.

\subsubsection{Cross-Strip Microchannel Plate}

The cross-strip MCP detector was built and optimized to meet the CHESS spectral resolution specifications at Sensor Sciences \citep{Vallerga+10,Siegmund+09}. The detector has a circular format and a diameter of 40 mm. The microchannel plates are lead silicate glass, containing an array of 10-micron diameter channels. They are coated with an opaque cesium iodide (CsI) photocathode, which provides QE = 15 $-$ 40\% at FUV wavelengths. When UV photons strike the photocathode to release photoelectrons, the photoelectrons are accelerated down the channels by an applied high voltage ($\sim$ 3100 V). Along the way, they collide with the walls of the channels, which produces a large gain over the initial single photoelectron. There are two MCPs arranged in a ``chevron" configuration. During flight, the detector achieved spatial resolution (PSF) of 25 $\mu$m over an 8k x 8k pixel format. The detector quantum efficiency (DQE) was measured by Sensor Sciences and is shown in Figure~\ref{fig:thruput} (left).

\subsubsection{Instrument Throughput and Effective Area}
The designed throughput of CHESS was set assuming that certain milestones in grating etching processes, specifically higher efficiency, low scattered light echelle grating, were achieved. CHESS was designed as a 2-bounce spectrograph to minimize the total reflective surfaces and maximum throughput of the instrument. The specified echelle diffraction efficiency into Littrow orders through the CHESS bandpass was defined at 70\%, a performance which matches echelle counterparts in visible and infrared spectrographs. The cross-strip MCP was coated with a photocathode to increase the DQE to $\sim$25\% average across the CHESS bandpass. The holographic ruling of the cross disperser has $\sim$50\% dispersion efficiency through the FUV. The reflective coating on both gratings is lithium fluoride over an aluminum layer (Al+LiF), which is $\sim$70\% reflective for $\lambda$ $>$ 1050 {\AA}. This puts the total throughput of the designed CHESS instrument at $\sim$4.3\% average across the FUV bandpass. With a total collecting area of 40 cm$^2$, the designed average effective area of CHESS was meant to be 1.75 cm$^2$.

\begin{figure} %[ht]
   \begin{center}
   \includegraphics[width=\textwidth]{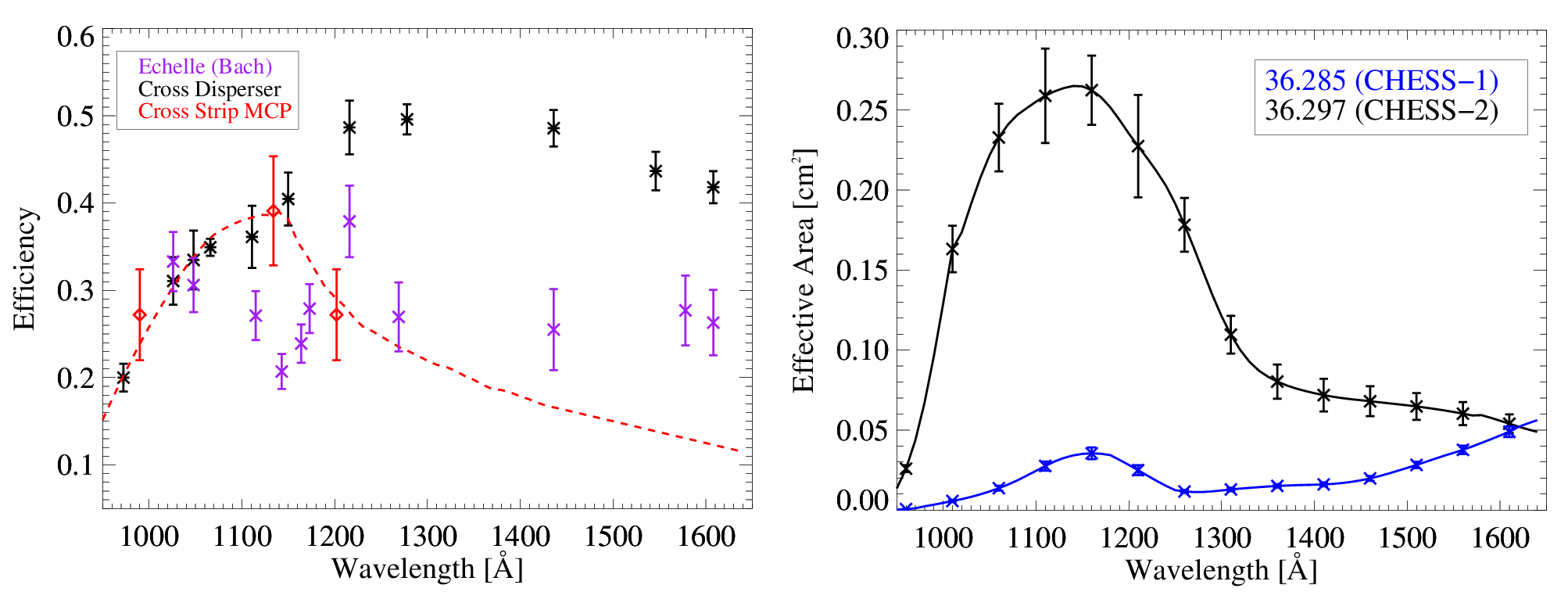}
   \end{center}
   \caption{\emph{Left:} Performance (for each grating: peak order efficiency, and for the detector: detector quantum efficiency) of all optical components of CHESS-2. For the first flight of CHESS, the only component performance that changed significantly was the echelle efficiency, which was $<<$0.1 (10\%) across the bandpass. \emph{Right:} The CHESS-2 effective area, including throughput loss from baffling, compared to the effective area of CHESS-1. The total effective area of CHESS-2 is about an order of magnitude larger than that of CHESS-1, owing primarily to the large gain in echelle order efficiency across the bandpass. Adapted from \citet{Hoadley+16}.} \label{fig:thruput} 
\end{figure}

The component-level performance and total throughput of CHESS-1 and CHESS-2, defined as the effective area (the total throughput of the instrument $\times$ 40 cm$^2$ collecting area), are shown in Figure~\ref{fig:thruput}. The low effective area of the instrument was driven by the efficiency of the echelle grating, which under-performed our specifications. The CHESS-2 effective area improved by roughly an order of magnitude from CHESS-1 by using a traditionally-ruled echelle grating, and subsequent flights of CHESS have improved the throughput by continuing to exchange echelle gratings \citep{Kruczek+17b}.

\begin{figure}
   \begin{center}
   \includegraphics[width=\textwidth]{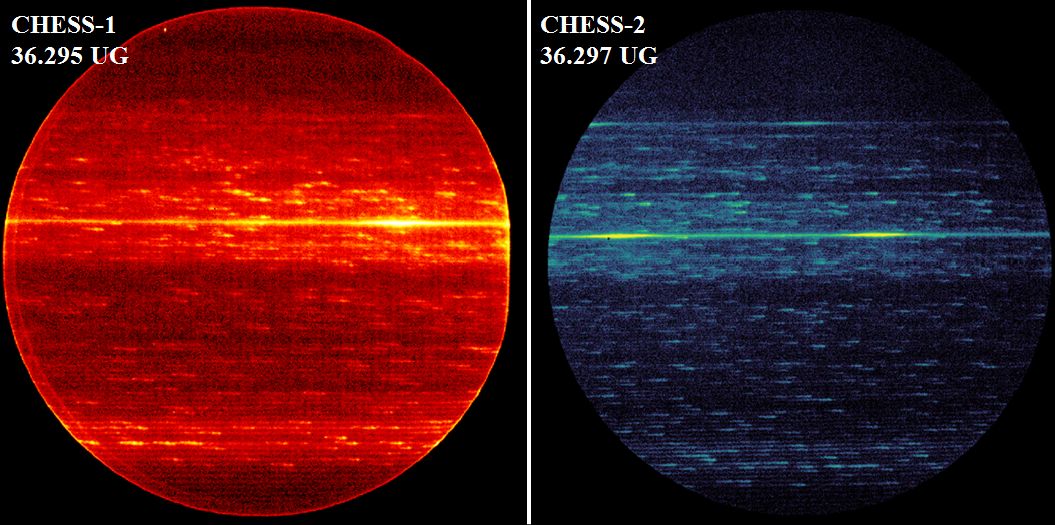}
   \end{center}
   \caption{Presented are the raw images of the CHESS-1 (left) and CHESS-2 (right; edge effects have been cropped out) echellograms from pre-flight calibrations (March 2014 and December 2015) using an arc lamp flowing 65\%/35\% H/Ar gas. The brightest feature in both images is H I-Ly$\alpha$ ($\lambda$ 1215.67 {\AA}); the CHESS-1 echellogram only shows Ly$\alpha$ in one echelle order, while the CHESS-2 echellogram disperses Ly$\alpha$ photons into two adjacent echelle orders. The other broad feature(s) visible in the CHESS-2 echellogram are HI-Ly$\beta$ (1025.72 {\AA}), about 1/4 of the way from the top of the image, and HI-Ly$\gamma$ (97.25 {\AA}), barely visible above the Ly$\beta$ features. The more discrete features dotted throughout the spectrum are H$_2$ emission from electron-impact fluorescence. Adapted from \citet{Hoadley+14} and \citet{Hoadley+16}.} 
   \label{fig:cals} 
\end{figure}

\section{Pre-flight Calibrations \& Data Reduction}\label{sec:reduction}

Pre- and post-flight calibrations of CHESS were performed in a dedicated vacuum chamber for sounding rocket payloads located in the Astrophysical Research Laboratory at the University of Colorado - Boulder. The chamber pumps the payload down to pressures of $<$ 10$^{-5}$ Torr, allowing for UV light to transmit through the payload and safe operation of electronics requiring high voltage, such as the MCP detector. Because CHESS was designed as a high-resolution spectrograph with a broad UV bandpass, a hollow-cathode arc lamp supplied with high purity H+Ar gas produced UV light used to focus and calibrate the instrument. This gas not only provided lines of hydrogen and argon, but a myriad of lines from H$_2$ fluorescence. Figure~\ref{fig:cals} shows the best-focused calibration images from CHESS-1 and CHESS-2 pre-flight spectra. The broadest emission features show atomic lines of hydrogen and argon, %with HI-Ly$\alpha$ at 1215.67 {\AA} being the brightest line in the entire image. The numerous, 
while the narrower features are H$_2$ fluorescence lines excited by interactions with energetic electrons within the lamp.

\begin{figure}
   \begin{center}
   \includegraphics[width=\textwidth]{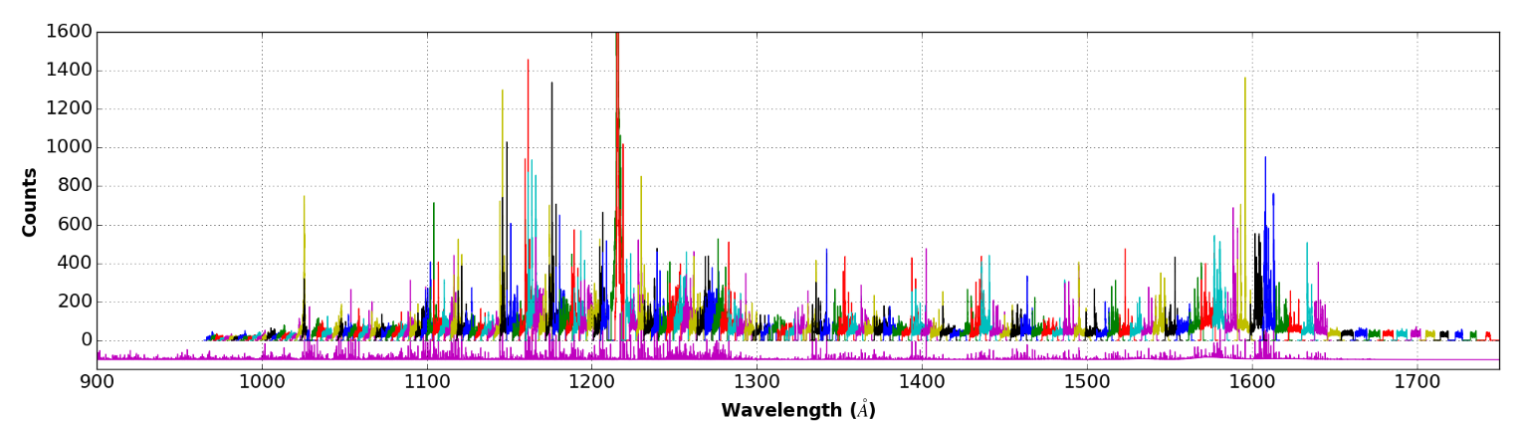}
   \end{center}
   \caption{The complete first-order wavelength solution for the pre-launch CHESS-2 calibration spectra from $\lambda \lambda$ 900 - 1750 {\AA}. The final wavelength solution was determined using H$_2$ fluorescence emission features and a functional extrapolation of the wavelength with a 6$^{th}$-order polynomial fit. Over-plotted in magenta is the model H$_2$ fluorescence inside the arc lamp (T$_{eff}$ = 800 K, N(H$_{2}$) $\sim$ 10$^{19}$ cm$^{-2}$, E$_{electron}$ = 50 eV). The spectrum is scaled to the highest total counts of the H$_2$ features; otherwise, Ly$\alpha$ would dominate the spectrum and the H$_2$ features would be washed out. To show how neighboring order spectra overlap and correlate to form the final 1D spectrum, individual order spectra have been plotted in different colors. Adapted from \citet{Hoadley+16}.}
   \label{fig:waves}
\end{figure}

Calibration images like those shown in Figure~\ref{fig:cals} were used to convert the CHESS echellogram to a 1-dimensional spectrum, defining both the wavelength solution and line spread function (LSF) of the instrument. First, each order dispersed by the cross disperser was identified and extracted to collapse into a 1D spectrum. Then, using adjacent order spectra, the orders were cross-correlated and stitched together to create a complete spectrum from 1000 - 1600 {\AA}. The cross-correlation was a (roughly) linear function from orders with shorter wavelengths to longer wavelength orders, so adjacent orders with no overlapping features could still be co-added into the complete spectrum. A synthetic H$_2$ model of electron-impact fluorescence was used to match H$_2$ features in the CHESS spectrum to the laboratory wavelengths of H$_2$ emission - this created the wavelength solution of the instrument, as shown in Figure~\ref{fig:waves}. 

\begin{figure}
   \begin{center}
   \includegraphics[width=1.0\textwidth]{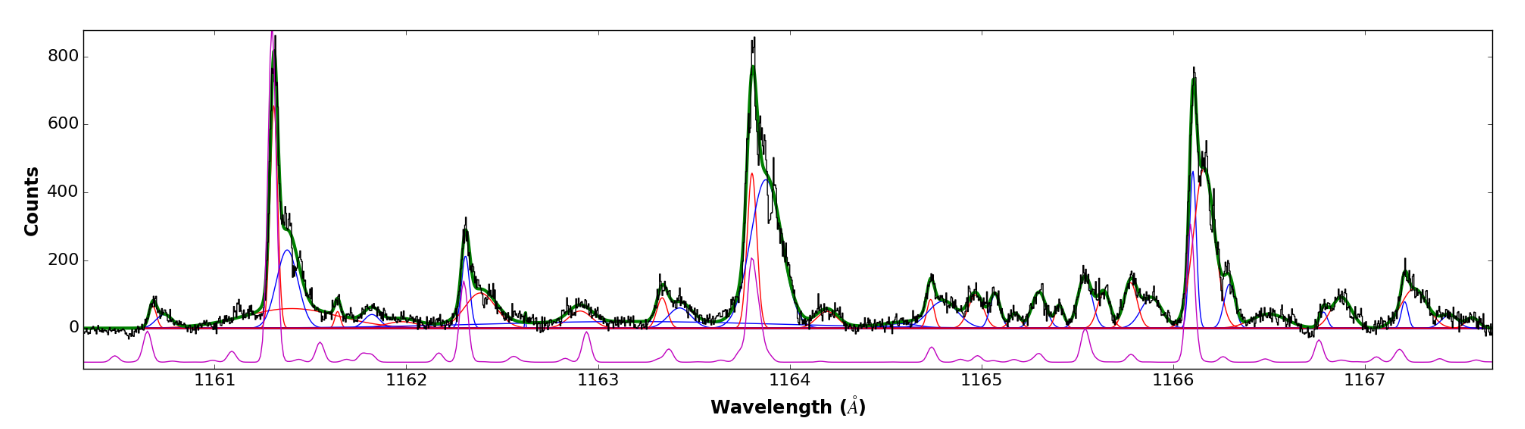}
   \end{center}
   \caption{The line spread function (LSF) fits of H$_2$ emission features in one order of the pre-launch calibration spectrum of CHESS-2 (echelle order m = 286). The order spectrum is shown in black. Red and blue Gaussian line fits are shown for the narrow and broad Gaussian fits for each line, respectively. The green line is the sum of all Gaussian components to reproduce the spectrum. A modeled H$_2$ fluorescence spectrum is shown in magenta. Adapted from \citet{Hoadley+16}.}
   \label{fig:lsf} 
\end{figure}

Once a wavelength solution was found, we convolved each synthetic emission line with a Gaussian of unknown width to match the CHESS emission line profiles, thereby creating LSF kernels through the CHESS bandpass. Each line was best fit with a narrow and broad Gaussian component. Figure~\ref{fig:lsf} shows a close-up of a few emission lines with a two-component Gaussian fit of the H$_2$ lines imaged by CHESS.

% figures here

The ``broad'' component of the CHESS LSF was unexpected and severely impeded the spectral resolution of the instrument. CHESS was designed with R $\geq$ 100,000, but the measured resolution was R $\leq$ 8,000. During the first build and calibration of CHESS, it was thought that the experimental echelle was the culprit of the degraded resolution, as it was etched onto a thin wafer that could bow or warp the surface of the grating. 

It was determined during the calibration of CHESS-2 that the broad emission lines were the result of the cross dispersing grating - the toroidal shape of the optic was designed to focus and correct for optical aberrations, but the ruling of the grating occurred along the optical axis perpendicular to the radius of curvature, which meant that, while the order widths were minimized to avoid order confusion, the spectral lines were out-of-focus. We verified this was the culprit by flipping the grating axis of the cross disperser in our Zemax models, which perfectly reproduced the line shapes and measured spectral resolution of the instrument \citep{Kruczek+18, Kruczek+2019}. While the cross disperser was never re-fabricated, \citealt{Kruczek+18} addressed the spectral resolution issue by adding a slight curvature to the high-dispersion echelle grating, achieving R $\sim$ 14,000 \citep{Kruczek+2019}. However, for CHESS-1 and CHESS-2, the spectral resolution of the instrument remained at R $<$ 8,000.

\section{CHESS-1: $\alpha$ Virgo}

The first flight of CHESS targeted the $\beta$ Scorpii$^1$ ($\beta$ Sco$^1$; HD 144217) sightline. $\beta$ Sco$^1$ is spectroscopic binary comprised of a B0.5V and a B1.5V spectral type star at distance of 161 pc with intermediate reddening (E(B-V) = 0.2, A$_V$ $\sim$ 0.6; log N(H$_2$) $\sim$ 19.8 cm$^{-2}$ \citep{Savage+77}). 

However, given the poor performance of the exerimental echelle grating used in CHESS-1, the science team elected to define a back-up target to move to if the signal-to-noise (S/N) of $\beta$ Sco$^1$ was insufficient to produce a science-quality spectrum during the limited time of the flight. The back-up target, $\alpha$ Virgo ($\alpha$ Vir; HD 116658; ``Spica''), is a spectroscopic binary, consisting of a B1III-IV star + B2V star at a distance of 43 pc \citep{Hoffleit+82}, well within the Local Bubble (E(B-V) = 0.03, A$_V$ $\sim$ 0.1 \citealt{Savage+77}). $\alpha$ Vir outputs 4 $-$ 5 times more far-UV flux than $\beta$ Sco$^1$ and thus demonstrated the capabilities of the CHESS instrument.

CHESS-1 (NASA/CU mission 36.285 UG) was launched from White Sands Missile Range on 24 May 2014 at 01:35am using a two-stage Terrier/Black Brant IX vehicle. Data was downlinked through the NASA telemetry system in real-time as [$x$, $y$, $t$, $PHD$] photon lists, where [$x$, $y$] defines the digital pixel position of the photon recorded on the detector, $t$ is the time the photon was recorded, and $PHD$ is the pulse height recorded for the photon, which is related to detector gain. Overall, the mission was a comprehensive success and achieved all the goals it aimed to meet. The instrument successfully collected data over the allotted $\sim$400 seconds of observing time. After the initial count rate on beta Sco was lower than expected, we elected to observe $\alpha$ Vir, and stayed on this target for the remainder of the flight. Count rate from $\alpha$ Vir was also lower than expected, but stellar absorption features in the echellogram made it clear that we were observing the star. Additionally, count rate contribution from air glow contamination was lower than expected, which suggested that at least one of the optics became unaligned with the instrument between pre-flight operations and launch. Post-launch calibrations verified that one of the gratings shifted before flight -- the echellogram did not appear on the detector until the instrument was shifted off-axis from the vacuum light source by $\sim$1 degree.

\begin{figure}
   \begin{center}
   \includegraphics[width=\textwidth]{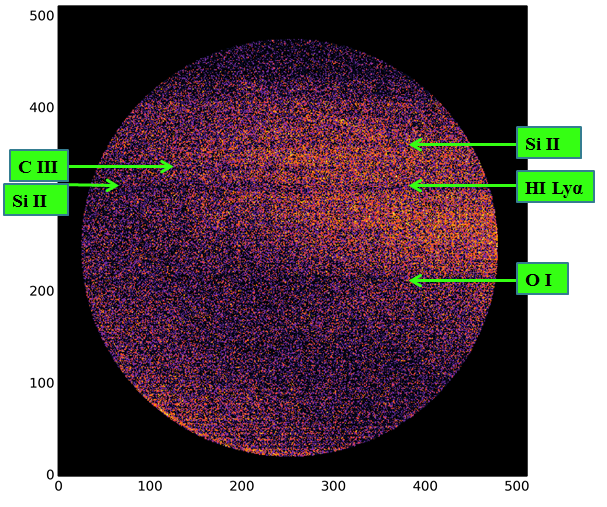}
   \end{center}
   \caption{The False-color representation of the flight echellogram from CHESS-1, taken on 24 May 2014, of $\alpha$ Vir. The purple/black regions represent areas with lower concentrations of photon counts, and blue/green pixels represent pixels with higher concentrations of photons collected. Marked with green arrows and labeled are the most prominent features in the echellogram. Because of the low S/N of the flight data, the echellogram has been binned to 512$\times$512, to show absorption features in the image. Adapted from \citet{Hoadley+14}.}
   \label{fig:chess1_flight}
\end{figure}

Figure~\ref{fig:chess1_flight} shows the echellogram of $\alpha$ Vir from the CHESS-1 launch. 
Prominent features apparent in the echellogram include stellar Ly$\alpha$, Si II (1193 {\AA}), Si III (1206 {\AA}), and C III (1175 {\AA}). However, the S/N of the flight data were not adequate to reconstruct an analysis-quality 1D spectrum.

\section{CHESS-2: $\epsilon$ Persei}

CHESS-2 was launched aboard a Terrier-Black Brant IX sounding rocket from White Sands Missile Range on 21 February 2016 as a part of NASA/CU mission 36.297 UG. 
CHESS-2 observed the line of sight to $\epsilon$ Persei ($\epsilon$ Per; HD 24760). $\epsilon$ Per is a B0.5III star  at d $\approx$ 300 pc with low$-$intermediate reddening (E(B-V) = 0.1; log(H$_2$) $\sim$ 19.5), indicating that the sightline may be sampling cool interstellar material. H$_{2}$, C I, CO, and C II were all detected by Copernicus; however, higher sensitivity and spectral resolution is required for a complete analysis of these types of sightlines \citep{Federman+80}. Observations by Copernicus and IUE have been used to measure the velocity structure along the sightline to $\epsilon$ Per, and have found at least three separate cloud structures described by different kinematic behavior and molecular abundances \citep{Bohlin+83}. Resolving the various molecular clouds on the $\epsilon$ Per sightline is the primary goal of CHESS-2. Overall, the line of sight to $\epsilon$ Per shows typical abundances of molecular material and ionized metal found in translucent clouds, such as H$_{2}$, Fe II and Mg II \citep{Bohlin+83}, consistent with the sightline towards recent star-forming sites.

\begin{figure}%[tbp]
	\begin{center}
	\includegraphics[width=0.95\textwidth]{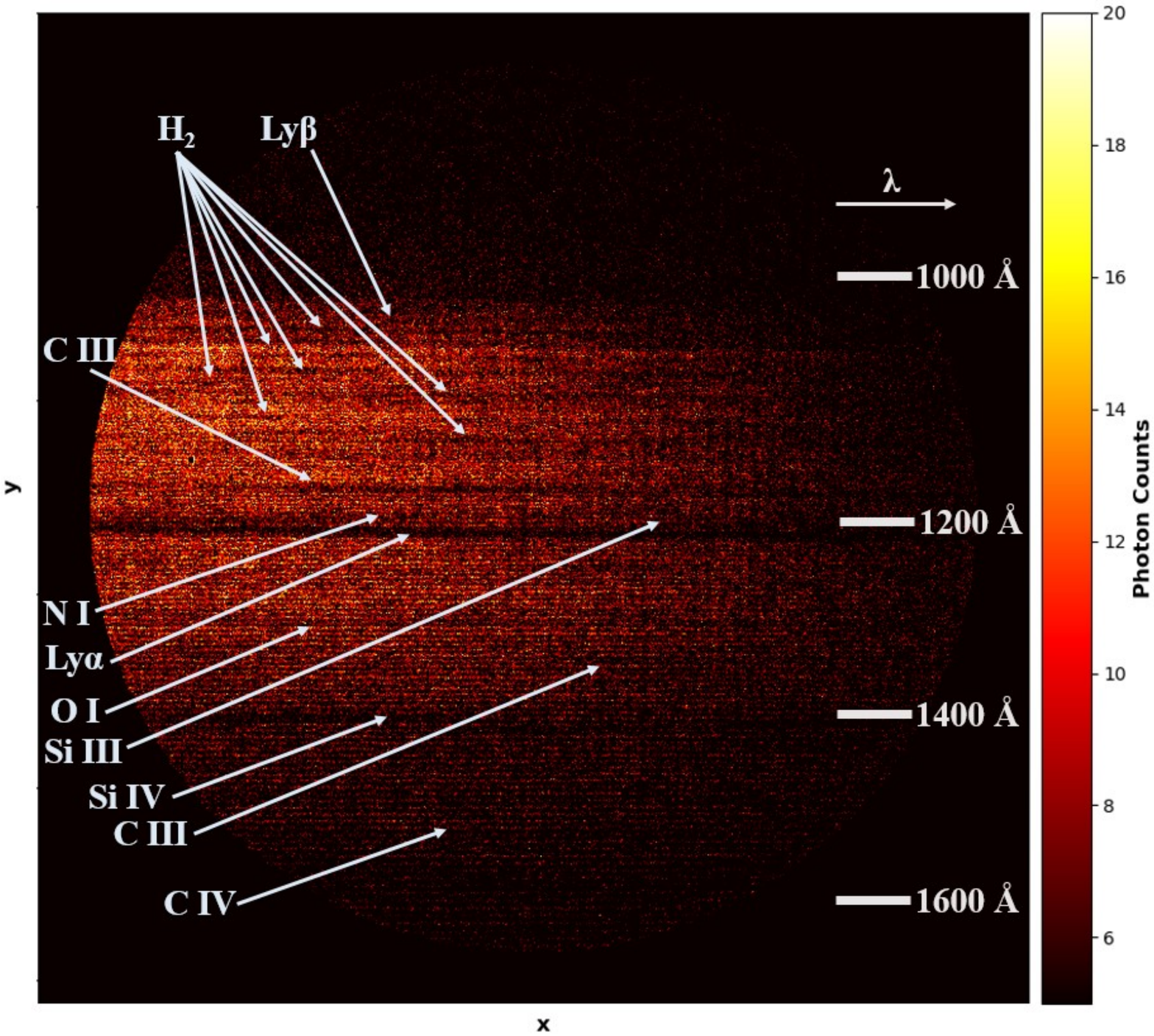} %, trim=0cm 0cm 0cm 6cm, angle=90
	\end{center}
	\caption{The raw, false-color echellogram of $\epsilon$ Per, recorded over t$_{exp}$ $\sim$ 250 sec on NASA/CU mission 36.297 UG (CHESS-2). Along the right, we mark the rough wavelength coverage as a function of the detector y axis. In an individual echelle order, wavelength increases to the right. Stellar continuum through the far-UV acts as a back-light behind the interstellar material. Dark streaks in the continuum show stellar and interstellar atoms, ions, and molecules absorbing photons at specific wavelengths; prominent interstellar and stellar features are labeled with arrows pointing to the absorption lines and the absorption species along the left. Adapted from \citet{Hoadley+16}.} \label{fig:chess2_raw}
\end{figure}

 The observed count rate on-target was $\sim$25,000 counts sec$^{-1}$ for a total exposure time (t$_{exp}$) of 250 seconds, resulting in a signal-to-noise ratio (S/N) per spectral resolution element of $\sim$ 5 from $\lambda$ = 1020 - 1060 {\AA} and S/N $\gtrsim$ 10 for $\lambda$ $>$ 1100 {\AA}. The CHESS-2 echellogram of $\epsilon$ Per is shown in Figure~\ref{fig:chess2_raw}.

\subsection{Analysis of CHESS-2 data}

The low S/N of the final flight data and diminished resolving power due to the fabrication error in the cross disperser meant that many of the primary science goals of CHESS were not achieved. We were unable to resolve any carbon-complex lines or achieve the signal needed to distinguish CO band absorption. However, the CHESS-2 spectrum of $\epsilon$ Per does show many absorption features between 1030 - 1220 {\AA}, which we fit and analyzed to compare to previous observations of this sightline taken with Copernicus and \emph{IUE}.

% CHESS-2: Raw Flight Echellogram

%\subsubsection{Scattered Light}

Scattered light from geo-coronal contamination posed a non-trivial amount of noise in the raw echellogram of CHESS-2. A combination of simulations and laboratory measurements were used to estimate the scattered light background in the raw data. Scattered light simulations were performed using the ray-tracing software Zemax using the average flux of geo-coronal HI-Ly$\alpha$ at night (63$\times$10$^{-15}$ erg cm$^{-2}$ s$^{-1}$ arcsec$^{-2}$), and scattered light images were taken using an uncollimated Bayard-Albert tube fed with hydrogen-argon gas \citep{France+06}. Both produced consistent scattered light profiles, which were used to model the background noise of the flight echellogram. We note that even after extracting the modeled scattered light profile, it appears that an underlying scattered light residual still remained, which appears as excess flux in absorption features through the 1D spectrum.  
For example, several known saturated absorption features are not entirely bottomed-out to the zero-point of flux scale. \emph{HST}-STIS suffers the same data reduction challenges due to the large scattered light profile of their mechanically-ruled gratings \citep{Landsman+97}. Initial data reduction packages for STIS produced similar results - for example, either the over- or under-subtraction of background light (e.g., \citealt{McGrath+99}). STIS handles the complicated echelle-dominated background light profile with a 2-D algorithm that still has issues completely getting rid of the scattered light noise (\citealt{Valenti+03} and references therein). 
This excess background noise in the CHESS observations is difficult to completely get rid of due to the low S/N of the observation, so we add this extra noise source into the errors in our analysis.

\begin{figure}%[htbp]
\centering
	\includegraphics[angle=90, width=\textwidth]{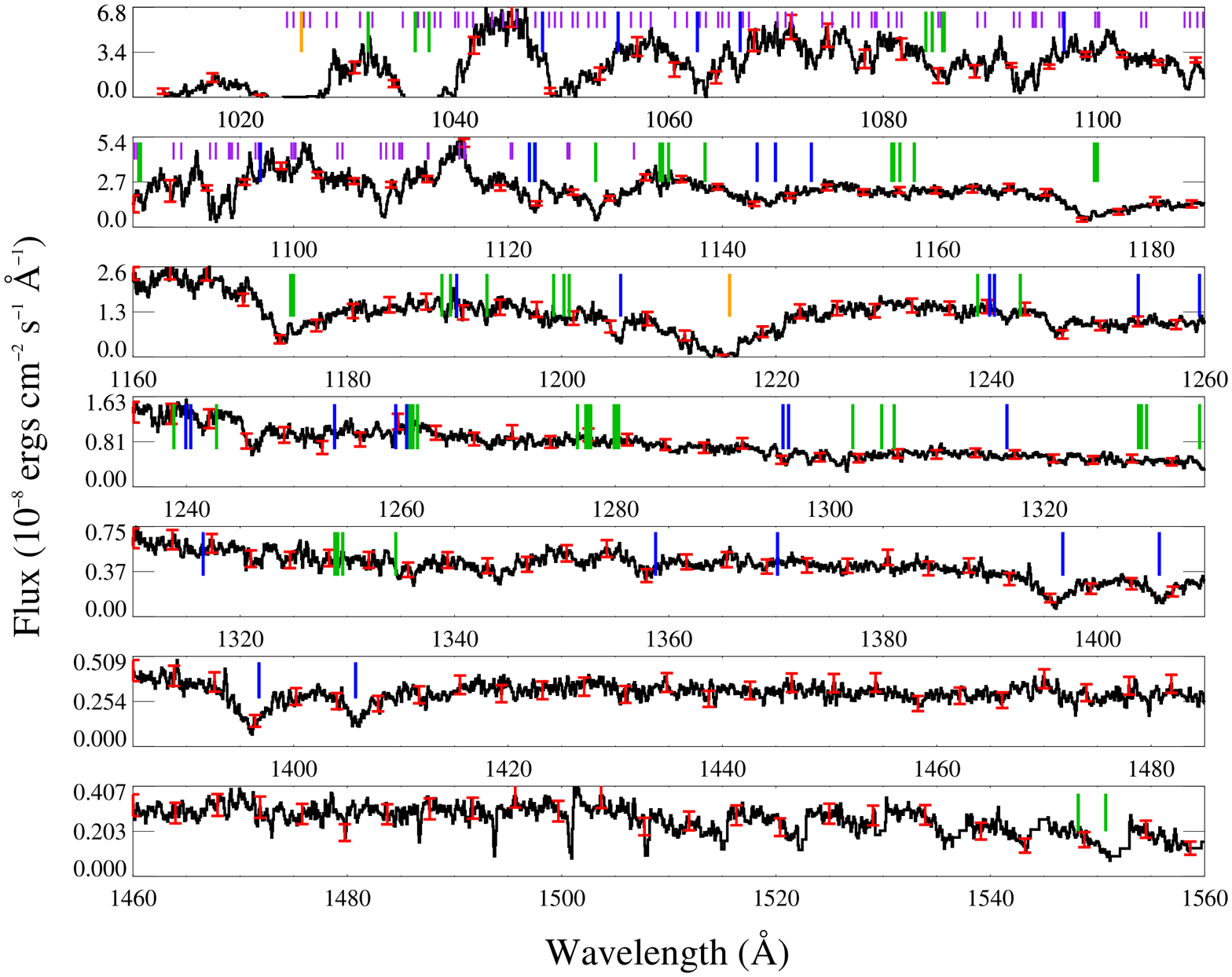}
	\caption{The flux-calibrated CHESS-2 spectra of $\epsilon$ Per from $\lambda$ = 1020 - 1550 {\AA}. Representative error bars are shown in red. The CHESS-2 spectrum was flux-calibrated against \emph{IUE} spectra of $\epsilon$ Per, and the continuum shape was compared with \emph{Copernicus} spectra at $\lambda$ $<$ 1150 {\AA}. Prominent stellar and interstellar absorption features are shown using vertical lines: purple lines show H$_2$ absorption features from $v$ = 0, $J$ = 0 - 7, orange lines show H I absorption, green lines show carbon, oxygen, and nitrogen species, and blue lines mark heavier metals (iron, silicon, sulfur, argon, magnesium, nickel, and copper).} \label{fig:spectrum}
\end{figure}

\subsubsection{Metal Absorption Lines}
%%
%% Metal absorption line figures
%%
%%
\begin{figure}
	\subfloat{\includegraphics[width=1.0\textwidth]{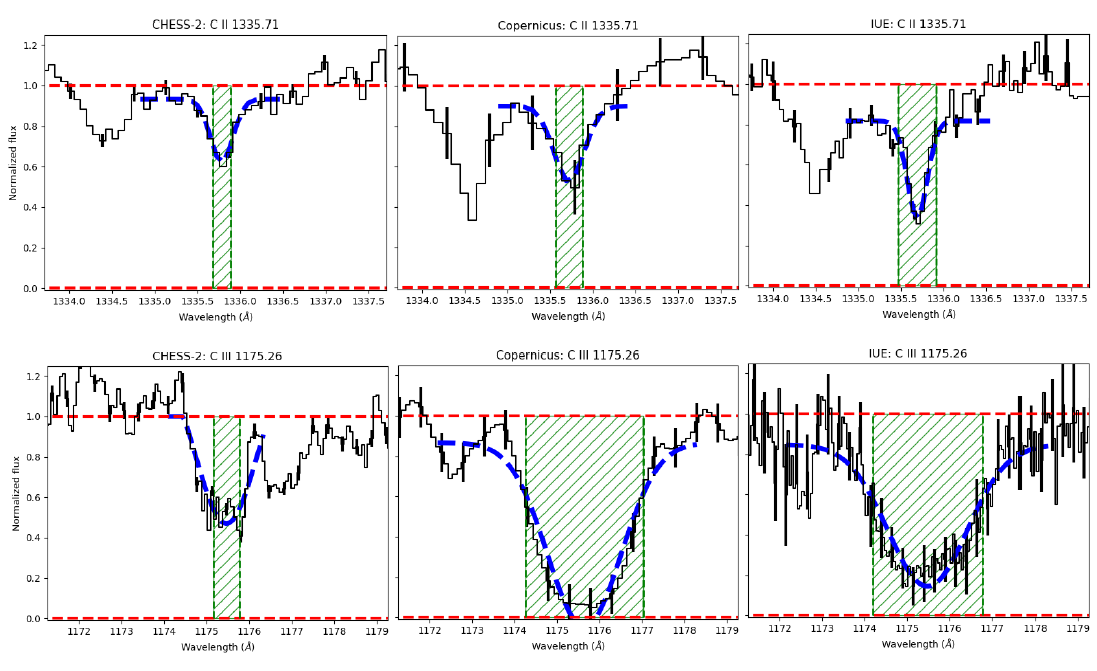}}
    \caption{(a) We present the normalized absorption features of different phases of interstellar carbon observed in the CHESS (left panels), Copernicus (middle panels), and \emph{IUE} (right panels) FUV spectra of $\epsilon$ Per. We fit all lines with a Gaussian profile (blue dashed line) and determine the equivalent width (W$_{\lambda}$) from the area under the Gaussian (green hashed area). We determine W$_{\lambda}$ for many metal lines found throughout the FUV in all data sets over a variety of interstellar phases - a comprehensive list of results is presented in Table~\ref{tab:metals}.} 
    \label{fig:metals}
\end{figure}
%%%%%%% Additional figures
\begin{figure}
\ContinuedFloat
    \subfloat{\includegraphics[width=1.0\textwidth]{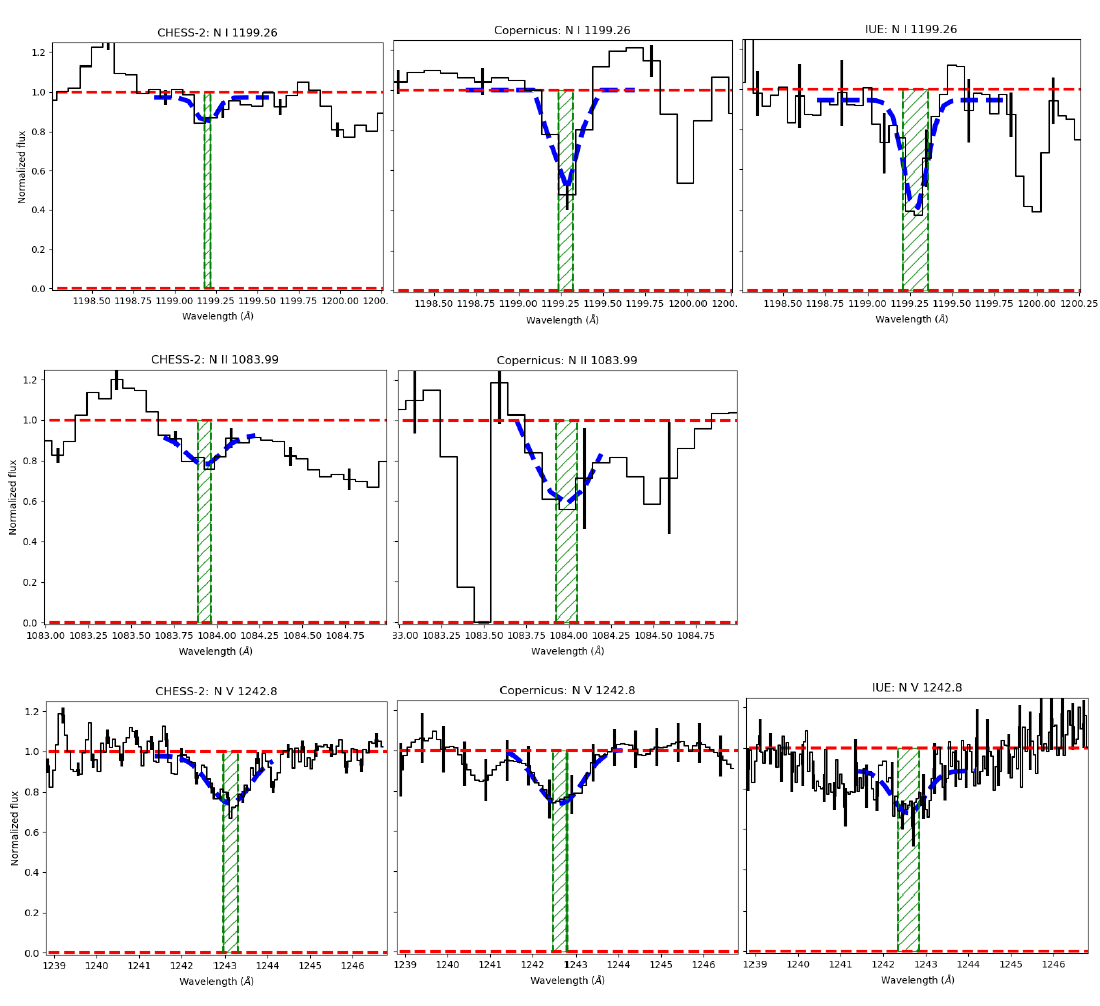}} 
    \caption{(b) Continued; different phases of nitrogen observed by CHESS, Copernicus, and \emph{IUE}, their Gaussian fits (blue dashed line), and W$_{\lambda}$ (green shaded area).}
\end{figure}
\begin{figure}
\ContinuedFloat
    \subfloat{\includegraphics[width=1.0\textwidth]{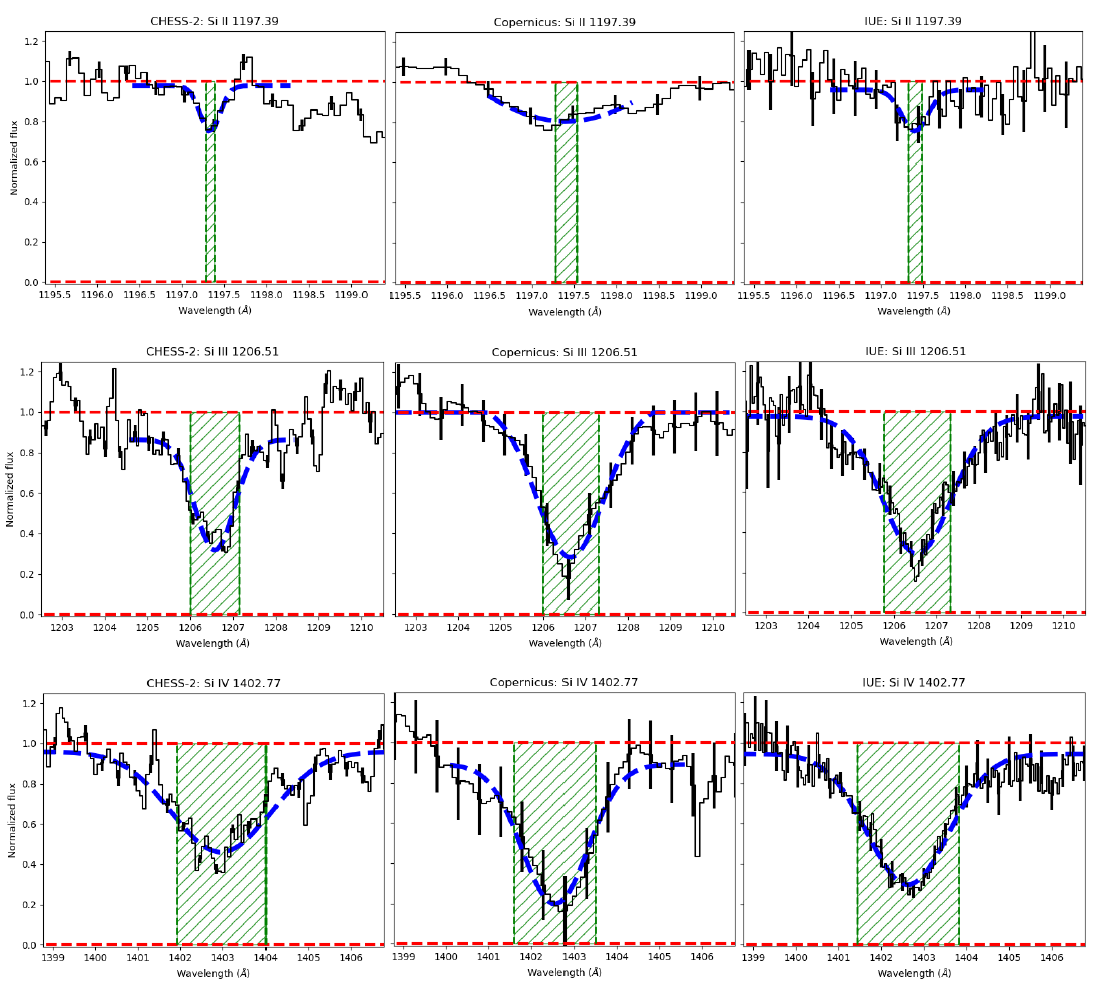}} 
    \caption{(c) Continued; different phases of silicon observed by CHESS, Copernicus, and \emph{IUE}, their Gaussian fits (blue dashed line), and W$_{\lambda}$ (green shaded area).}
\end{figure}
\begin{figure}
\ContinuedFloat
    \subfloat{\includegraphics[width=1.0\textwidth]{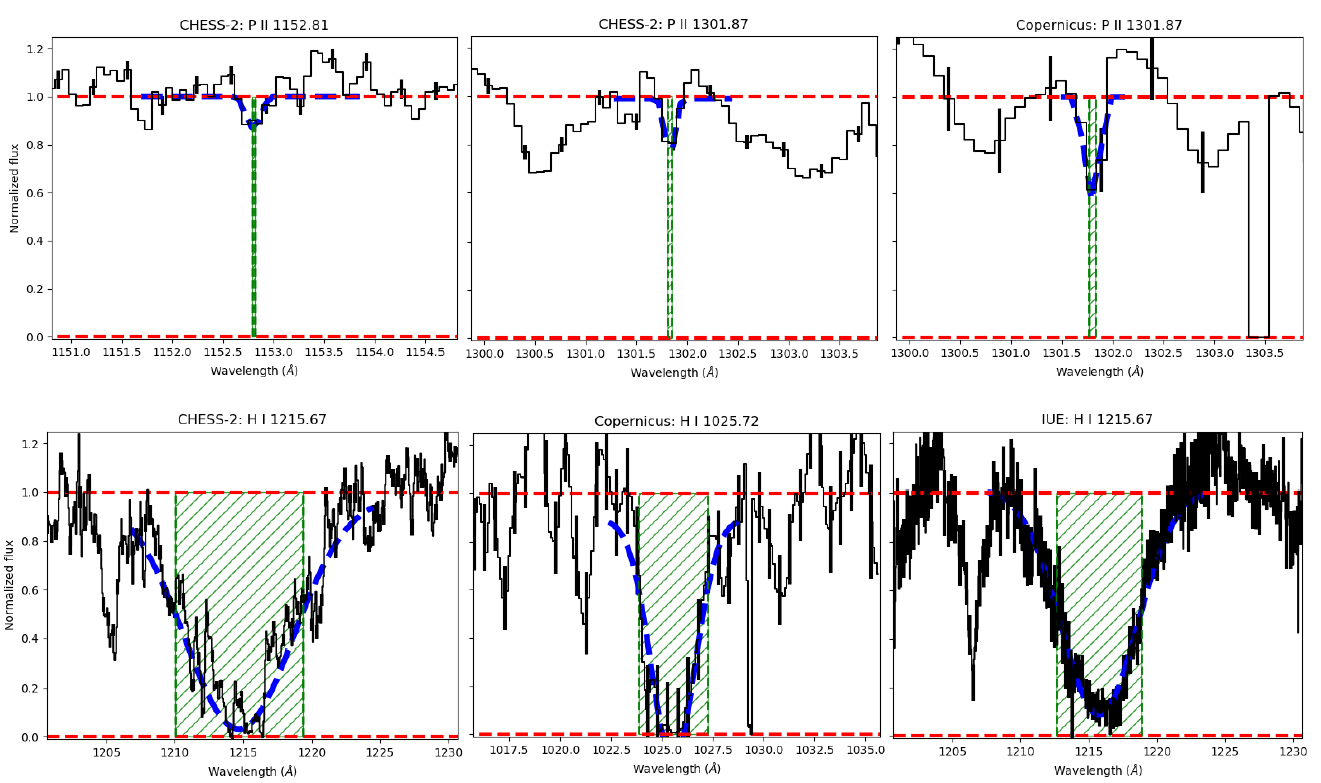}} \\
    \subfloat{\includegraphics[width=1.0\textwidth]{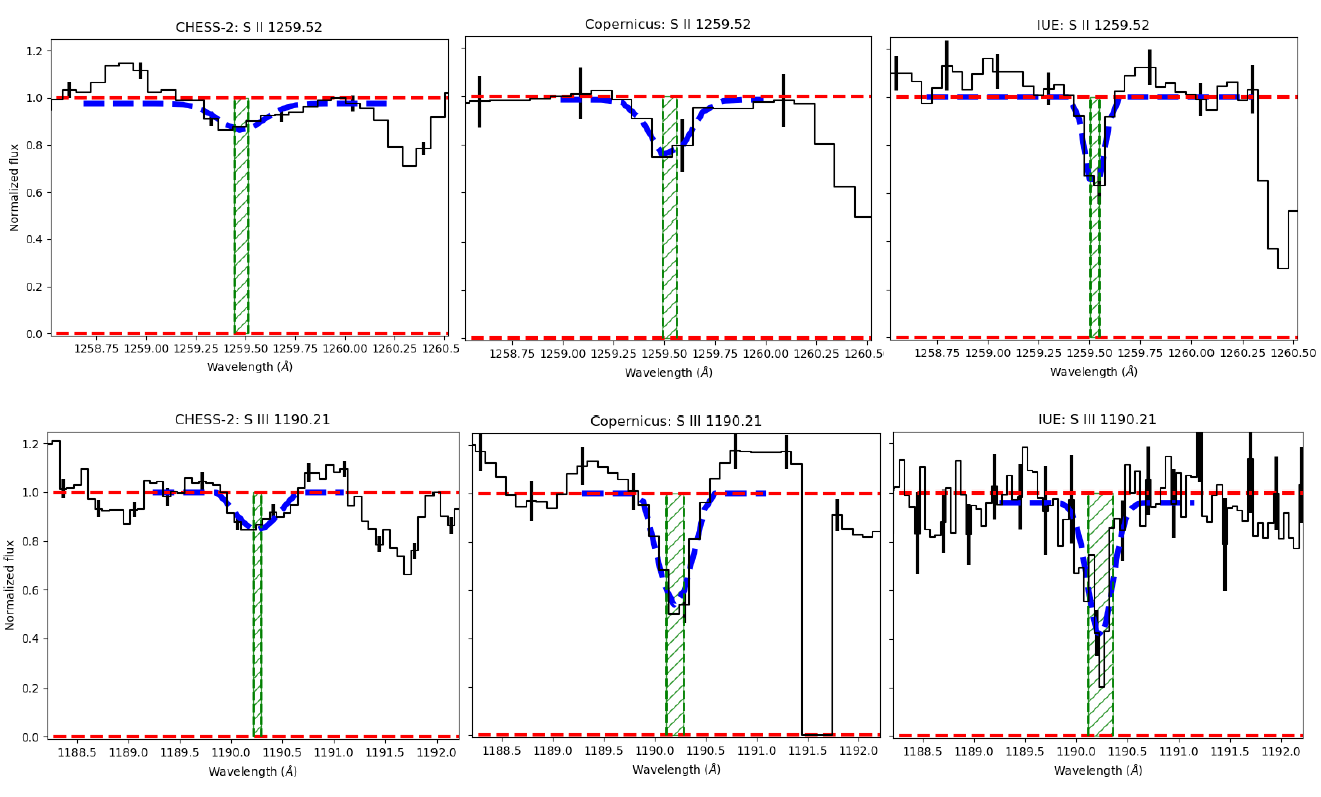}}
    \caption{(d) Continued; phosphorus, hydrogen, and sulfur observed by CHESS, Copernicus, and \emph{IUE}, their Gaussian fits (blue dashed line), and W$_{\lambda}$ (green shaded area).}
\end{figure}

\begin{table}
\caption{$\epsilon$ Persei Metal Absorption Line Diagnostics {\&} Comparison with Archival \emph{IUE} and Copernicus Spectra} 
\label{tab:metals}
\begin{center}       
\begin{tabular}{l|c|c|ccc|ccc|ccc} 
       \hline \hline
				&						&			&		&		CHESS-2				&						&			&	\emph{IUE}	&							&		&Copernicus	&		\\
		Ion	&	$\lambda$	&	$f$	&	FWHM	&	W$_{\lambda}$	&	$\chi ^2$	& FWHM	&	W$_{\lambda}$	&	$\chi ^2$	& FWHM	&	W$_{\lambda}$	&	$\chi ^2$	\\
				&	({\AA})		&			&	(km/s)	&	(m{\AA})			&					&	(km/s)	& (m{\AA})	&							&	(km/s)	&	(m{\AA})	&	\\
    \hline
	H I	&	1025.72	&	0.079	&	1011    &	3646.6	&	156.60	&				&						&			&	743	&		3373.2	&	17.00	\\
	H I	&	1215.67	&	0.416	&	1835	&	7701.1	&	247.06	&	1620	&		6194.6	&	54.26	&			&						&				\\

	C II	&	1334.53	&	0.129	&	131	&		190.8	&	1.21	&	99	&		347.6	&	0.51	&	124	&		360.0	&	1.13	\\
	C II	&	1335.71	&	0.115	&	69	&		208.6	&	1.25	&	63	&		448.6	&	0.99	&	93	&		316.9	&	1.35	\\
	C III	&	1175.26	&	0.272	&	440	&		624.1	&	31.73	&	561	&		2569.3	&	13.12	&	521	&		2763.5	&	11.8	\\
	
	N I	&	1134.98	& 0.040	&	75	&		35.1	&	3.04	&			&					&				&	60	&		95.5	&	1.15			\\
	N I	&	1199.26	& 0.133	&	37	&		36.3	&	3.27	&	40	&		150.5	&	7.67	&	53	&		84.7	&	6.23			\\
	N II	&	1083.99	& 0.101	&	65	&	74.0	&	2.03	&			&					&				&	157	&		121.9	&	1.16	\\
	N V	&	1238.82	& 0.156	&	113	&		209.3	&	12.12	&	283	&		339.6	&	10.28	&	325	&		407.9	&	5.27	\\
	N V	&	1242.80	&	0.078	&	270	&		330.6	&	12.32	&	220	&		509.6	&	13.37	&	307	&		322.7	&	7.28	\\
	
	O I	&	1302.17	&	0.052	&	218	&		302.8	&	15.59	&	160	&		430.1	&	10.41	&	242	&		467.2	&	50.29			\\
	
	S II	&	1259.52	&	0.016	&	64	&		69.0	&	2.22	&	28	&		42.6	&	2.23	&	52	&		66.6	&	1.07		\\
	
	Si II	&	1194.50	&	1.62	&	96	&		122.9	&	3.11	&	113	&		348.5	&	14.34	&	303	&		267.6	&	5.34			\\
	Si II	&	1197.39	&	0.323	&	69	&		107.9	&	12.13	&	92	&		161.8	&	17.18	&	101	&		260.7	&	8.03			\\
	Si III	&	1206.51	&	1.67	&	273	&		1151.2	&	2.94	&	473	&		1558.8	&	10.62	&	470	&		1337.1	&	5.78		\\
	Si IV	&	1393.76	&	0.528	&	570	&		1999.4	&	9.48		&	648	&		2698.0	&	16.71	&	637	&		2673.1	&	9.78		\\
	Si IV	&	1402.77	&	0.262	&	583	&		2077.3	&	5.68		&	481	&		2387.4	&	12.94	&	385	&		1906.1	&	5.15		\\\
	
	P II	&	1152.81	&	0.236	&	51	&		22.1	&	0.89	&			&					&				&			&					&					\\
	P II	&	1301.87	&	0.017	&	37	&		38.4	&	2.06	&	45	&		37.5	&	0.04	&	50	&		68.2	&	0.48		\\
	
	\hline
\end{tabular}
\end{center}
\end{table}

We visually inspect the CHESS-2 spectrum of $\epsilon$ Per against the \emph{IUE} and Copernicus spectra of the sightline. %The low S/N of the CHESS flight data make it difficult to confidently point out absorption features against the stellar spectrum, except for very dark absorption features (e.g., H$_2$ complexes, HI 1026 and 1216 {\AA}, SiIV 1402, etc). 
From this inspection, we select a number of stellar and interstellar absorption features that match well between all data sets. Next, we normalize the continuum of all data sets around the absorption lines of interest by fitting an arbitrary quadratic function through continuum points surrounding the absorption feature. Each absorption feature is fit with a Gaussian line profile using a reduced-$\chi^2$ statistic. The initial conditions of the Gaussian fit are found to not drastically change the final result, so all line profiles begin with the same width parameter ($\sigma$). We then find the equivalent width ($W_{\lambda}$) of the each absorption line by finding the area under each Gaussian profile. This entire process is repeated for the same absorption lines found in \emph{IUE} and Copernicus spectra.

Overall, we find good agreement between all three data sets. Figure~\ref{fig:metals} presents the normalized absorption lines for a large selection of metal feature from all three data sets, and Table~\ref{tab:metals} describes the quantities derived from each line profile, including the full width half maximum (FWHM) of each absorption line fit and $W_{\lambda}$ derived from the line fits. 
%For shallow lines, and assuming that the optical depth ($\tau$) $<$ 1, we can estimate a total column density of the species in the sightline ($N(X)$) by assuming the lines fit along the linear part of the curve of growth. We find good agreement in our estimates of $N(X)$ and \citet{Martin+82} for lines that overlap. For example, for CII 1334, we find log$_{10}N($C II$)$ = 16.3 from the CHESS-2 spectrum and 16.6 from the \emph{IUE} and Copernicus data sets (and log$_{10}N($C II 1335$)$ = 16.4 for CHESS-2 and 16.7 for \emph{IUE}/Copernicus), whereas \citet{Martin+82} determine 16.9 in the densest cloud. 
Table~\ref{tab:nx} presents column density estimates from CHESS-2 and our re-analysis of \emph{IUE} and Copernicus spectra of $\epsilon$ Per, along with the total column density of each species determined by \citet{Martin+82} when available. We estimate a total column density of the species in the sightline ($N(X)$) by assuming the lines fit along the linear part of the curve of growth. We find good agreement in our estimates of $N(X)$ and \citet{Martin+82} for all lines that overlap. For saturated lines (e.g., H I), we use the intrinsic properties of the line to estimate the column density in the sightline, which is the same approach used for our H$_2$ analysis below (see Section~\ref{sec:abs} for more details). We discuss the H I column density estimate from the CHESS-2 data in the following subsection.

\begin{table}
\caption{$\epsilon$ Persei Metal Column Densities} 
\label{tab:nx}
\begin{center}       
\begin{tabular}{l|ccc|c} 
    \hline \hline
    	    	&		    &	log$_{10}N(X)$	&				&	log$_{10}N(X)$ (total)	\\
        Ion		&	CHESS	&	\emph{IUE}     	&   Copernicus  &   \citet{Martin+82}	\\
    \hline
	C II	&	16.3$^{+0.3}_{-0.1}$	&	16.6$^{+0.1}_{-0.1}$	&	16.6$^{+0.2}_{-0.1}$ 	& 16.9	\\
	
	N I		&	16.3$^{+0.0}_{-0.6}$	&	16.3$^{+0.0}_{-0.1}$	&	16.4$^{+0.3}_{-0.3}$	&	16.3	\\
	N II	&	16.3$^{+0.1}_{-0.2}$	&		&	16.5$^{+0.2}_{-0.3}$	&	16.0	\\
	
	O I	&	17.0$^{+0.2}_{-0.1}$	&	17.1$^{+0.1}_{-0.1}$	&	17.0$^{+0.1}_{-0.2}$	&	17.2	\\
	
	S II	&	16.9$^{+0.2}_{-0.5}$	&	16.6$^{+0.1}_{-0.4}$	&	16.8$^{+0.4}_{-0.4}$	&	16.3		\\
	
	Si II	&	15.4$^{+0.4}_{-0.3}$	&	15.6$^{+0.3}_{-0.0}$	&	15.8$^{+0.4}_{-0.3}$	&		\\
	
	P II	&	13.6$^{+0.2}_{-0.1}$	&	13.5$^{+0.1}_{-0.1}$	&	13.8$^{+0.0}_{-0.3}$	&	13.8		\\

	\hline
\multicolumn{5}{l}{$^{\star}$ \citet{Jenkins+79}.}
\end{tabular}
\end{center}
\end{table}

\subsubsection{H$_2$ Absorption Profile Fitting, Rotation Diagrams, and Formation Rates} \label{sec:abs}

Molecular hydrogen (H$_2$) is the most abundant constituent of interstellar clouds. Interstellar H$_2$ is observed as absorption features throughout the far-UV regime along sightlines toward hot stars. Many H$_2$ interstellar sightlines display multiple temperature components \citep{Spitzer+73}, where the low rotation ($J$) lines ($J$ = 0, 1) provide a measure of the kinetic (collision-dominated) gas temperature, while H$_2$ probed in higher rotational levels ($J$ $>$ 3) are sensitive to other physical processes, like UV-pumping (e.g., \citealt{vanDishoeck+86}), formation on dust grains (e.g., \citealt{Jura+74,Lacour+05}) and heating by turbulence and/or shocks (e.g. \citealt{Gry+02,Ingalls+11}). The multi-component population structure provides important diagnostics about which processes dominate the physical cloud conditions and individual interstellar cloud properties, such as the total hydrogen density and the H$_2$ formation rate. 

To analyze the properties of interstellar H$_2$ in the $\epsilon$ Per sightline, we determine the H$_2$ rotational structure following the procedure of \citealt{France+13a}. We fit a multi-component H$_2$ absorption model to observed absorption features by combining the \texttt{H$_2$ools} optical depth templates \citep{McCandliss+03} and the \texttt{MPFIT} least-squares minimization routine \citep{Markwardt+09}. This method takes the theoretical line shape of each H$_2$ energy level for a given set of column density (N(H$_2$)) and Doppler-$b$ values, convolves the synthetic spectrum with the line spread function (LSF) of the instrument, and adjusts N(H$_2$) and $b$ until a best-fit spectrum is found. 

The $\epsilon$ Per spectrum is normalized to our model spectra around select H$_2$ absorption bands. We fit synthetic profiles simultaneously to the H$_2$(1 $-$ 0) ($\lambda_0$ 1092.2 {\AA}) and H$_2$(0 $-$ 0) ($\lambda_0$ 1108.1 {\AA}) complexes for low-to-intermediate rotational levels ($J^{\prime \prime}$ = 0 $-$ 7). These bands are uncontaminated by other stellar and interstellar absorption features and have the highest S/N H$_2$ absorption features in our data set. To check the solutions of the H$_2$(1 $-$ 0) and H$_2$(0 $-$ 0) profile fits, we fit the modeled H$_2$ column densities to the H$_2$(4 $-$ 0) H$_2$ band ($\lambda_0$ 1049.4 {\AA}), which is also free of stellar and interstellar contaminants. 

While it has been shown that the $\epsilon$ Per sightline may have upwards of four interstellar cloud components \citep{Martin+82}, the moderate spectral resolution of CHESS does not allow us to separate these different components. Therefore, assuming a single H$_2$ cloud, we find a $b$-value of 3.6 km s$^{-1}$, determined from the H$_2$($J^{\prime \prime}$ = 2 $-$ 6) rotational levels, which are sensitive to changes in $b$. This value is consistent with previous curve-of-growth measurements of H$_2$ in the $\epsilon$ Per sightline (e.g., \citealt{Stecher+67,Carruthers+71}) and typical $b$-values for H$_2$ in the local ISM \citep{Lehner+03,France+13a}.

\begin{figure}
	\centering
	\includegraphics[angle=90,width=0.49\textwidth]{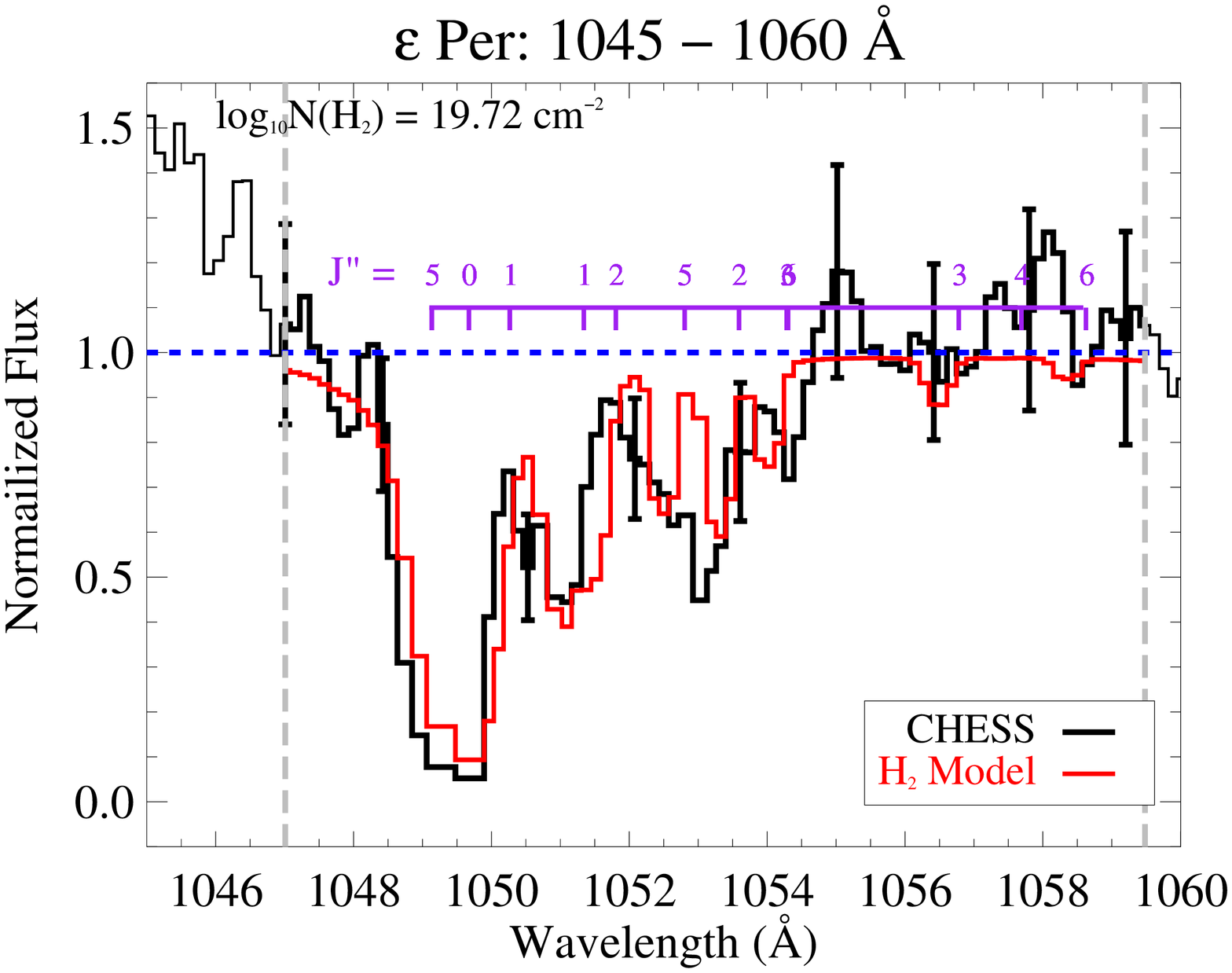}
    \includegraphics[angle=90,width=0.49\textwidth]{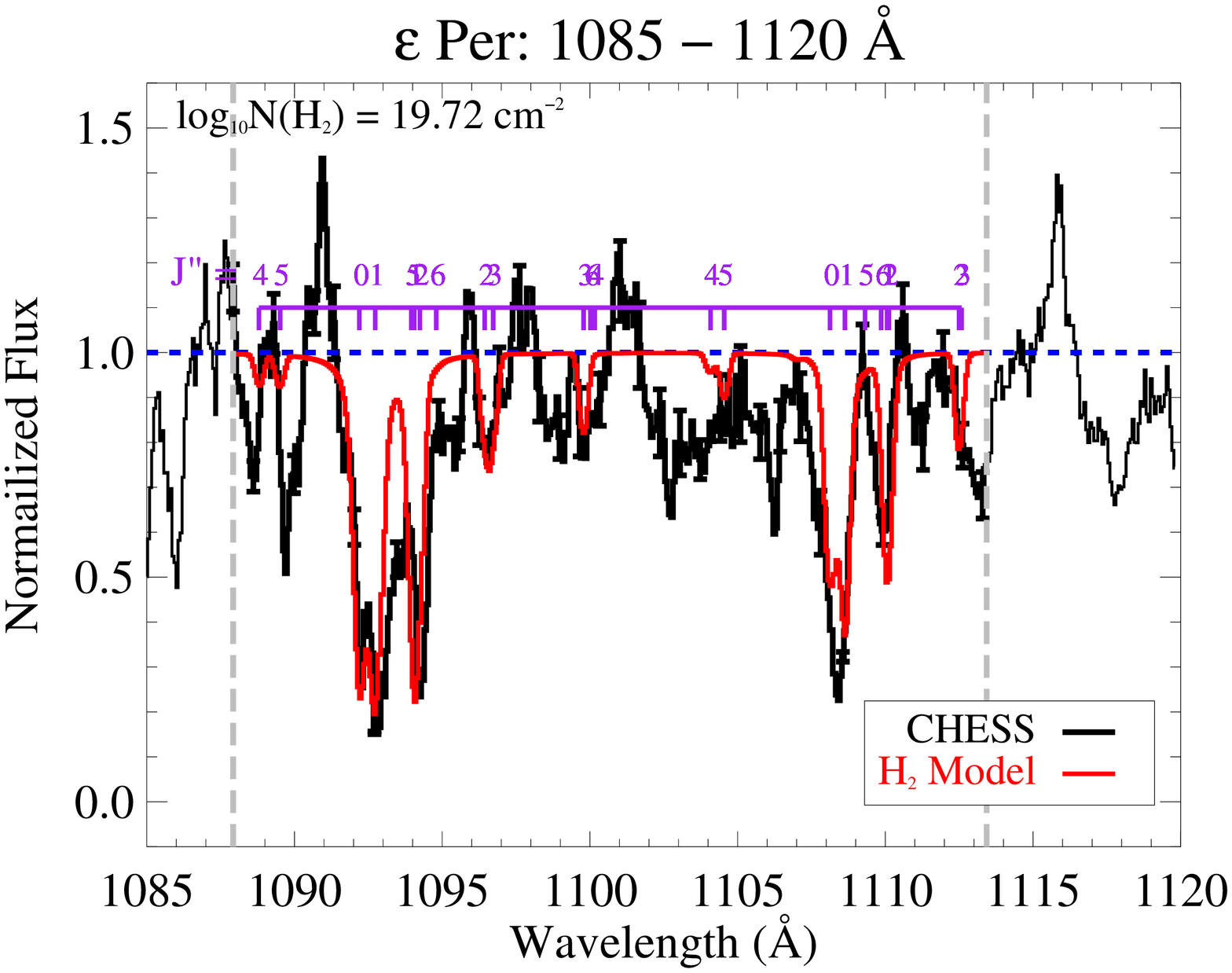} %\\
    \caption{Synthetic H$_2$ profile fits for the H$_2$(4 $-$ 0) band (\textit{left}) and H$_2$(1 $-$ 0) and H$_2$(0 $-$ 0) bands (\textit{right}), shown in red, are overlaid on top of the $\epsilon$ Per spectrum. Molecular rotational levels are labeled with purple dashes. The best-fit Doppler velocities for all three spectral band fits is $b$ = 3.6 km s$^{-1}$.} \label{fig:epsPer_H2profs}
\end{figure}

\begin{table}
\caption{$\epsilon$ Persei H$_2$ Parameters from CHESS-2 Observations and Model Fits} 
\label{tab1}
\begin{center}       
\begin{tabular}{l|c|c} 
    \hline \hline
    H$_2$ Level & log$_{10}$N(H$_2$,$v''$=0,$J''$) & log$_{10}$N(H$_2$,$v''$=0,$J''$) \\
                &           MPFIT Model            &            MCMC Model            \\
    \hline
	$J''$ = 0	& 19.20$^{+0.31}_{-0.02}$ 	&  19.65         \\
	$J''$ = 1	& 19.56$^{+0.42}_{-0.25}$	&  19.52           \\
	$J''$ = 2	& 17.35 $\pm$ 0.41			&  17.47            \\
	$J''$ = 3	& 15.47$^{+0.40}_{-0.32}$	&  15.66           	\\
	$J''$ = 4	& 14.75$^{+0.68}_{-0.38}$	&  14.62        	\\
	$J''$ = 5	& 14.91$^{+0.51}_{-0.48}$   &  14.61      		\\
	$J''$ = 6	& 13.12$^{+0.50}_{-0.42}$   &  13.51          		\\
	$J''$ = 7	& $<$ 13.51 			    &  13.40         		\\
	\hline \hline
	log$_{10}N$(H$_{2}$)	&	19.72 $\pm$ 0.35    & 19.83 $\pm$ 0.02          \\
	log$_{10}N$(\ion{H}{1})	&	20.31 $\pm$ 0.19    & 20.43 $\pm$ 0.07         \\
	T$_{01}$	&  95 $\pm$ 2 K                     & 95 $\pm$ 1 K         \\
	T$_{exc}$	&  500 $\pm$ 150 K                  &          \\
	$b_{H_2}$	&  3.6 $\pm$ 0.6 km s$^{-1}$        & 3.6 km s$^{-1}$         \\
	$f(H_2)$ 	&  0.24 $\pm$ 0.05                  & 0.30 $\pm$ 0.03         \\
    $Rn$  	    &                                   & 49.5 $\times$ 10$^{-16}$ s$^{-1}$         \\
	$n_H$   	&                                   & 55 cm$^{-3}$      \\
	\hline
\end{tabular}
\end{center}
\end{table}

We present the total H$_2$ column density and densities in each $J$-level in Table~\ref{tab1}. The normalized spectra and best-fit models for the H$_2$(4 $-$ 0) and H$_2$(1 $-$ 0)+H$_2$(0 $-$ 0) bands of $\epsilon$ Per are shown in Figure~\ref{fig:epsPer_H2profs}. We determine the total column density of \ion{H}{1} ($N$(\ion{H}{1})$_{obs}$) to be log$_{10}N$(\ion{H}{1})$_{obs}$ = 20.33 $\pm$ 0.19 from the Ly$\alpha$ absorption feature ($\lambda$ 1215.67 {\AA}) using the same procedure as the H$_2$ absorption fits (listed in Table~\ref{tab1}). However, as noted in \citet{Diplas+94}, for B-stars, there is a non-negligible stellar contribution to the \ion{H}{1}-Ly$\alpha$ absorption profile, such that $N$(\ion{H}{1})$_{obs}$ = $N$(\ion{H}{1})$_{stellar}$ + $N$(\ion{H}{1})$_{interstellar}$. For $\epsilon$ Per, \citet{Diplas+94} found the total stellar \ion{H}{1} column density to be log$_{10}N$(\ion{H}{1})$_{stellar}$ = 18.82, so the total \emph{interstellar} column density of neutral hydrogen is log$_{10}N$(\ion{H}{1}) = 20.31 $\pm$ 0.19. We use $N$(\ion{H}{1}) and $N$(H$_2$) to calculate the molecular fraction ($f(H_2)$) of the interstellar cloud:
\begin{equation}
	f(H_2) = \frac{2 N(H_2)}{N(HI) + 2N(H_2)} = 0.24 \pm 0.05.
\end{equation}
The homonuclear nature of the hydrogen molecule forbids radiative transitions from $J^{\prime \prime}$ = 1 $\rightarrow$ 0 within the same electronic band, while quadrupole transitions from $J^{\prime \prime}$ = 2 $\rightarrow$ 0 are allowed but slow ($A_{2 \rightarrow 0}$ $\approx$ 3 $\times$ 10$^{-11}$ s$^{-1}$; \citealt{Wolniewicz+98}). Therefore, for a sightline with an appreciable density of hydrogen through the ISM, collisions are expected to control the level populations of the lower energy $J^{\prime \prime}$ states (0, 1, 2). We can therefore define a kinetic temperature, $T_{01}$, which describes the collision-dominated regime of the interstellar cloud. $T_{01}$ is derived from the ratio of column densities in the $J^{\prime \prime}$ = 0 and $J^{\prime \prime}$ = 1 levels:
\begin{equation}
	N(J^{\prime \prime} = 1)/N(J^{\prime \prime} = 0) = \frac{g_1}{g_0} e^{\left( -E_{01}/kT_{01} \right)} = 9 e^{\left( -171K/T_{01} \right)}
\end{equation}
\noindent
where $g_1$ and $g_0$ are the statistical weights in the $J^{\prime \prime}$ = 1 and 0 levels, respectively. We find that the kinetic temperature of the H$_2$ in the $\epsilon$ Per sightline is $T_{01}$ = 95 $\pm$ 2 K.

\begin{figure}%[htbp]
\centering
	\includegraphics[angle=90,width=0.49\textwidth]{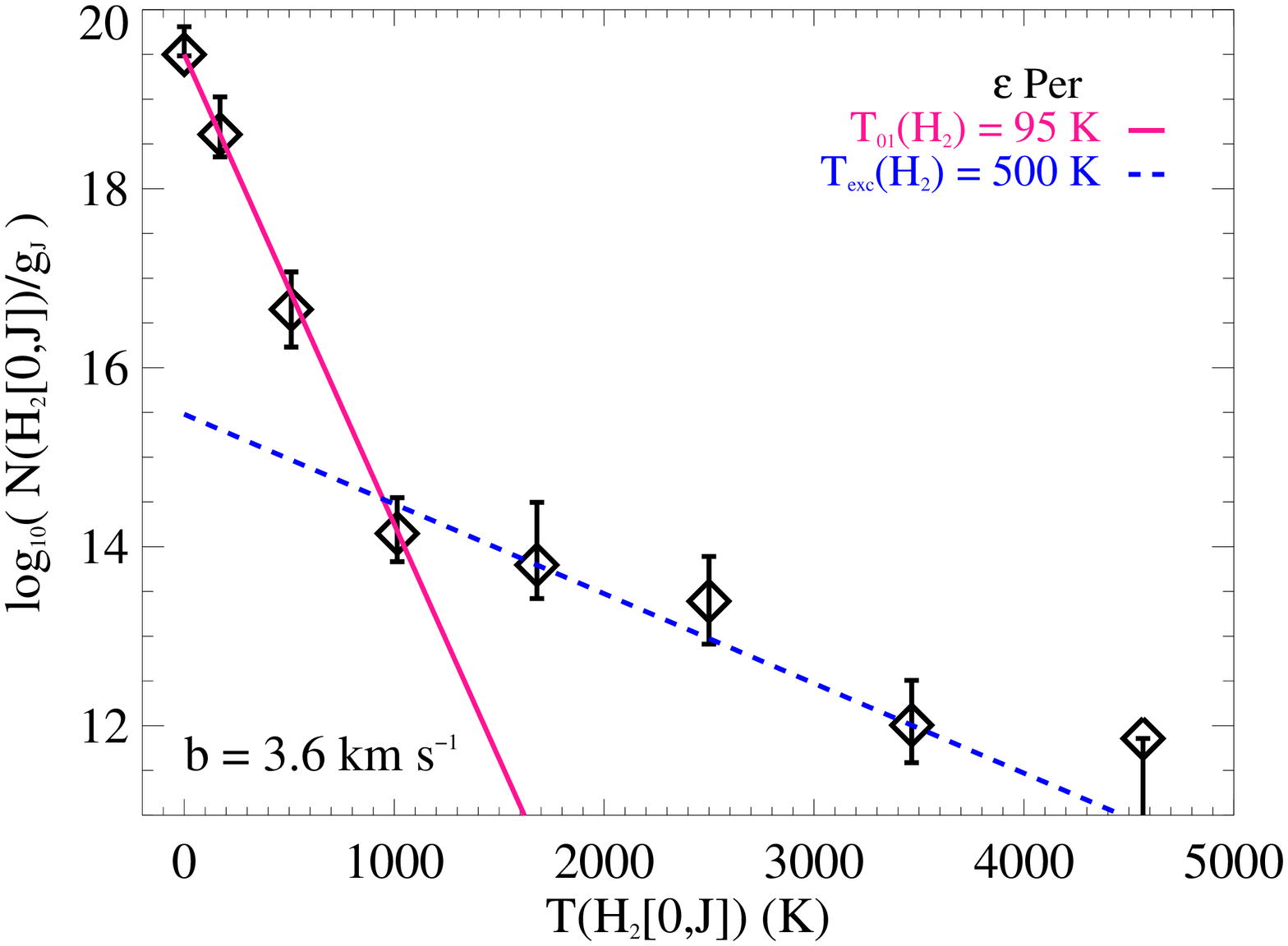} %, trim=0cm 0cm 0cm 6cm, angle=90
	\includegraphics[angle=90,width=0.49\textwidth]{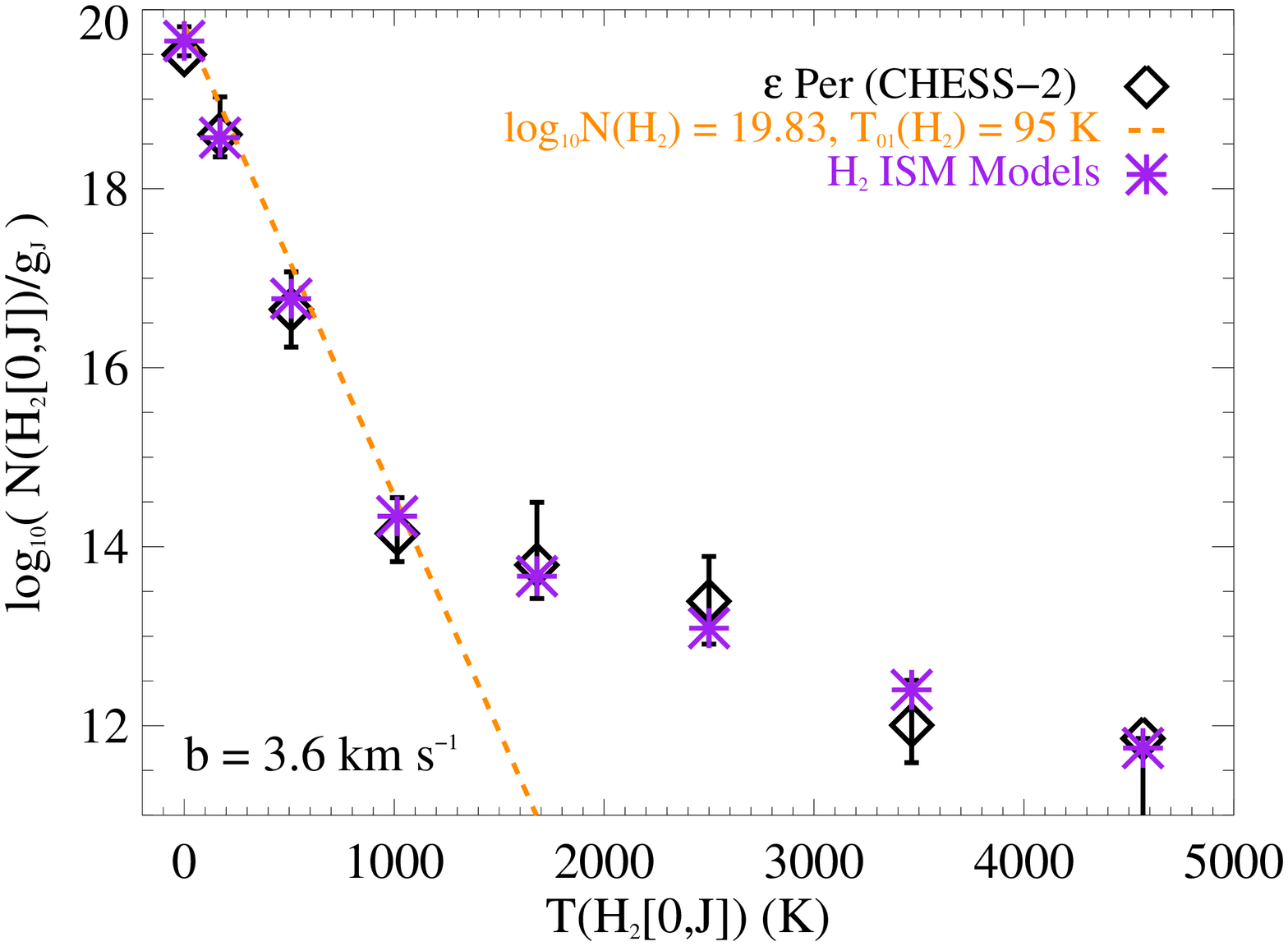}
	\caption{We present the $\epsilon$ Per H$_2$ rotation diagram with two different model fits. \emph{Left:} The H$_2$ rotation diagram is fit assuming the sightline has two temperature populations of H$_2$: a cool, kinetic temperature, described by $T_{01}$ (pink), and a warmer, ``excitation'' temperature, described by $T_{exc}$ (blue). \emph{Right:} The H$_2$ rotation diagram is fit with an H$_2$ equilibrium model (purple asterisks), which includes affects from UV-photon pumping, collisions with other particles, and formation/destruction rates of H$_2$ in a diffuse medium. The kinetic temperature derived from these models is shown in orange.} \label{fig:models}
\end{figure}

The temperature of the higher rotational levels ($J^{\prime \prime}$ $>$ 3) can be fit with an ``excitation'' temperature, $T_{exc}$, by determining the slope of the rotation diagram between $J^{\prime \prime}$ = 3 $-$ 7. A least-squares linear fit through these rotation levels found $T_{exc}$ = 500 $\pm$ 150 K (Figure~\ref{fig:models}, left).

We use the new results from the CHESS-2 observations to measure these two interstellar physical cloud parameters in the $\epsilon$ Per sightline. First, we derive $Rn$ and $n_H$ from two empirical approaches: the first from \citet{Martin+82}, who use an updated prescription to a common approach (e.g., \citealt{Jura+75a}) to define diffuse ISM physical cloud conditions, and the second from \citet{Gry+02}, who take into account additional physical mechanisms important for denser clouds, like dust and molecular shielding and irradiation. The \citet{Martin+82} analysis calculates $Rn$ based on the formalism presented in \citet{Jura+75a}: $Rn$~$\approx$~$N$(H$_2$;~J=4)~/~$N$(\ion{H}{1})(4.28$\times$10$^{8}$). Using the column densities derived from the CHESS-2 observation, we find $Rn$ = 45.6$\times$10$^{-16}$ s$^{-1}$. The \citet{Gry+02} method balances formation and destruction rates of H$_2$: $Rn$~=~0.5$\frac{f}{1-f}~\beta_0~\langle~S~\rangle$, where $f$ is the molecular fraction of H$_2$ in the interstellar cloud, $\beta_0$ is the mean rate of H$_2$ photo-dissociation in the diffuse local interstellar medium, and $\langle S \rangle$ is the mean self-shielding factor of H$_2$ against photo-dissociation from the UV radiation source. For the diffuse ISM, $\beta_0$ is defined as 5.0$\times$10$^{-11}$ s$^{-1}$ (e.g., \citealt{Habing+68,Mathis+83}). We find $\langle S \rangle$
$\approx$ 8.2$\times$10$^{-5}$ using Equation 37 in \citet{Draine+96}, which depends on $N$(H$_2$) and $b$ derived from our synthetic absorption spectra (Section~\ref{sec:abs}). We find $Rn$ $\approx$ 7.5$\times$10$^{-16}$ s$^{-1}$ using the \citet{Gry+02} approach, which is $\sim$6 times lower than $Rn$ derived using the \citet{Martin+82} method. Table~\ref{tab:Rn} presents the H$_2$ formation rates and resulting realizations of $n_H$ using $R$ found from \citet{Jura+75a}, \citet{Black+87}, and \citet{Draine+11}.

\begin{table}

\caption{Physical Cloud Conditions of $\epsilon$ Persei, Derived from Analytic Solutions}
\label{tab:Rn}
\begin{center}       
\begin{tabular}{l|l|l} 
\hline  \hline
Method: & \citet{Martin+82}	&	\citet{Gry+02} \\
\hline
$Rn$ (s$^{-1}$)	&	45.6 $\times$ 10$^{-16}$	&	6.6 $\times$ 10$^{-16}$ \\
$n_H^{a}$ (cm$^{-3}$) &	150		&	20			\\
$n_H^{b}$ (cm$^{-3}$) &	80		&	10			\\
$n_H^{c}$ (cm$^{-3}$) &	10		&	1.5			\\
\hline
\end{tabular}
\end{center}
\begin{tabular}{l}
$^{a}$ $R$ = 3.0 $\times$ 10$^{-17}$ cm$^3$ s$^{-1}$; \citet{Jura+75a}. \\
$^{b}$ $R$ = 6.0 $\times$ 10$^{-18}$ T$^{1/2}$ cm$^3$ s$^{-1}$, assuming T = T$_{01}$ = 95 K; \citet{Black+87}. \\
$^{c}$ $R$ = 7.3 $\times$ 10$^{-17}$ (T / 100)$^{1/2}$ $\langle \epsilon_{gr} \rangle$ $\Sigma_{-21}$ cm$^3$ s$^{-1}$, assuming T = T$_{01}$ = 95 K, $\Sigma _{-21}$ = \\
\qquad the grain surface area for silicate-graphite PAHs = 6.0 ($\times$ 10$^{-21}$) cm$^2$ H$^{-1}$, and $\langle \epsilon_{gr} \rangle$ \\
\qquad = the average efficiency for H$_2$ formation on dust grains $\sim$ 1.0, which gives a lower limit \\
\qquad to $n_H$; \citet{Weingartner+01, Draine+11}.
\end{tabular}
\end{table}

There are roughly two dex in spread for $n_H$ between these two methods, depending on the adopted H$_2$ formation rate. To better constrain $Rn$ and $n_H$ in the $\epsilon$ Per sightline, we create stationary H$_2$ equilibrium models, which populate H$_2$ energy levels based on the physical conditions of an interstellar cloud. 
We follow the framework presented by \citet{Draine+96} (Equations 1 - 14), \citet{Gry+02}, and references therein. The models balance excitation and de-excitation processes working on H$_2$, including formation pumping, H$_2$ dissociation, photo-excitation, spontaneous decay, and collisional de-excitation.

The UV radiation field is assumed to be dominated by the blackbody UV continuum produced by $\epsilon$ Per. Interstellar H$_2$ is excited to higher energy electronic levels by this UV radiation, where it either fluoresces back to the ground electronic energy state or dissociates. Energy levels, transition probabilities, and dissociation probabilities for fluorescent transitions are adapted from \citet{Abgrall+89}, \citet{Abgrall+92, Abgrall+93a}, and \citet{Abgrall+93b}. Spontaneous decay rates of ground-level quadrupole transitions are taken from \citet{Wolniewicz+98}. Finally, collisional de-excitation of H$_2$ is calculated from equations and coefficients presented by \citet{Mandy+93} and \citet{Martin+95}, where hydrogen is the dominate collisional partner of H$_2$. 

We allow $Rn$ and $n_H$ to float but constrain the values of $n_H$ between the minimum and maximum values presented in Table~\ref{tab:Rn}. We assume that $n$(H), $n$(H$_2$), and $n_H$ are uniformly distributed across the interstellar cloud slab, which allows us to retrieve $N$(H$_2$) and $N$(H I) from our simulations, and $Rn$ does not vary across the modeled region. The models are run through a Markov Chain Monte Carlo (MCMC) routine, performed with the Python \texttt{emcee} package \citep{Foreman+12} in the same manner as described in \citet{Hoadley+17}. The routine is run with 100 walkers over 500 individual steps. The MCMC analysis produces comparable molecular and atomic hydrogen column densities compared to our direct observational results: log$_{10}N$(H$_2$) = 19.83 $\pm$ 0.02 and log$_{10}N$(H I) = 20.43 $\pm$ 0.07, with the same kinetic temperature: T$_{01}$(H$_2$) = 95 $\pm$ 1 K. The formation rate and hydrogen density in the physical cloud model are found to be $Rn$ = 49.5 $\times$ 10$^{-16}$ s$^{-1}$ and $n_H$ = 55 cm$^{-3}$. Our results are slightly larger values than diffuse cloud conditions of $\epsilon$ Per determined by \citet{Jura+75b} and \citet{Martin+82} and much larger than the denser interstellar cloud schematic of \citet{Gry+02}, making our results consistent with a diffuse interstellar cloud origin \citep{Jura+75a}. Our results find that the volumetric formation rate of H$_2$ in the $\epsilon$ Per sightline is $R$ = 9.0 $\times$ 10$^{-17}$ cm$^3$ s$^{-1}$. Assuming the gas density is uniform, we find that the interstellar cloud probed in the sightline extends $\sim$0.4 pc. 

We show the best-fit H$_2$ equilibrium models, along with the CHESS-2 H$_2$ rotation diagram for $\epsilon$ Per, in Figure~\ref{fig:models} (right). Our resulting H$_2$ formation rate is slightly higher than the classic formation rates presented by \citet{Jura+75a} and \citet{Black+87} (Table~\ref{tab:Rn}). This can be accounted for in a few ways: either the efficiency of H$_2$ formation on silicate-dominated dust grains is higher than expected ($\langle \epsilon_{gr} \rangle$ $\sim$ 20\%), or the distribution of dust grain sizes in the $\epsilon$ Per sightline deviates from the canonical distribution assumed for the average Milky Way ISM \citep{Weingartner+01}.

\section{Conclusions}
% Future flights: CHESS-3 done and publication soon based on same/similar modeling techniques (CITE Kruczek et al. 2017?). CHESS-4 target and launch date mentions here.
The Colorado High-resolution Echelle Stellar Spectrograph (CHESS) is a UV spectroscopic instrument meant to demonstrate high-throughput, high-resolution far-UV spectroscopy. Such concepts are especially important for pathfinders of upcoming space missions, where future observatories will have to address key observational capabilities when \emph{HST} is no longer available. In addition, CHESS provides an ideal platform to improve and test experimental technologies meant to vastly improve the performance of diffraction gratings, mirror coatings, and detector efficiencies throughout the UV.

For the first two launches of CHESS, we tested and flew different high-order diffraction gratings (echelles) for UV spectroscopy and demonstrated the improved performance and dynamic range of state-of-the-art MCP detectors. In the end, both the underwhelming performance of the experimental echelle gratings and an error in the ruling of the toroidal surface figure cross-dispersing grating proved detrimental to the overall performance of the spectrograph. Still, the first two launches of the instrument gathered data on the stellar sightlines they observed, with the second flight acquiring adequate S/N to perform a detailed analysis of the interstellar sightline. We find many different metallic absorption lines in the $\epsilon$ Per sightline, and we fit multiple H$_2$ band complexes with thermal and stationary equilibrium models to characterize the H$_2$ in this diffuse ISM sightline. We find the column density and temperature of cool H$_2$ to be consistent with analyses performed with Copernicus and find that the full stationary equilibrium model suitably fits the observed column densities of all $J$ states.

The CHESS experiment went on to observe nearby early B-type stars to study the molecular properties of local interstellar clouds aboard two additional NASA-funded sounding rocket missions: the third launch of CHESS took place at WSMR in June 2017 and successfully observed the sightline towards $\beta$ Sco$^1$ \citep{Kruczek+17b}, while the fourth and final launch took off from the Kwajalein Test Range in April 2018 and observed $\gamma$ Arae \citep{Kruczek+18}. A detailed analysis of both sightlines, paired with an analysis of archival spectral datasets obtained with Copernicus and {\it FUSE}, has shown that previous studies overestimate the average gas kinetic temperature of the diffuse molecular ISM by 12\% \citep{Kruczek+2019}.

%\acknowledgments
We acknowledge the hard work and dedication of the NASA WFF/NSROC payload team, the Physical Sciences Laboratory at New Mexico State University, and the Navy team at WSMR that supported the NASA/CU 36.297 UG. We thank the referee of this manuscript for their helpful suggestions and points where further clarification was warranted. KH would like to acknowledge the generous support and guidance from Prof. Jim Green, Ted Schultz, Michael Kaiser, and the University of Colorado UV sounding rocket research group during the build and operations of CHESS-I and CHESS-II. KH would also like to thank Dr. Nicholas Kruczek, Jacob Wilson, Jack Swanson, and Nicholas Erickson for each of their individual contributions and achievements that led to a successful CHESS-II flight, subsequent CHESS successes, and their invaluable moral support before, during, and after the mission. KH acknowledges support by the David {\&} Ellen Lee Postdoctoral Fellowship in Experiment Physics at Caltech. This research was funded by the NASA Astrophysics Research and Analysis (APRA) grant NNX13AF55G. AY acknowledges support by an appointment to the NASA Postdoctoral Program at Goddard Space Flight Center, administered by the Universities Space Research Association through a contract with NASA.

%\bibliography{h2_bibliography}

\end{document}